\documentclass[lettersize, journal]{IEEEtran}

\usepackage{float}
\usepackage{caption}
\usepackage{color}

\usepackage{cite}
\usepackage{amsmath,amssymb,amsfonts}
\usepackage{amsthm}
\usepackage{graphicx}
\usepackage{textcomp}
\usepackage{xcolor}
\usepackage{svg}
\usepackage{subfigure}
\usepackage{mathtools}

\usepackage{algpseudocode}
\usepackage{siunitx}
\usepackage{subcaption}
\usepackage{booktabs}
\usepackage{multirow}
\usepackage{algorithm}
\usepackage{algpseudocode}

\newtheorem{lemma}{Lemma}

\allowdisplaybreaks
\makeatletter
\def\ScaleIfNeeded{%
\ifdim\Gin@nat@width>\linewidth \linewidth \else \Gin@nat@width
\fi } \makeatother

\graphicspath{{Fig/}}

\begin{document}

\title{Continuous Aperture Array (CAPA)-Based Secure Wireless Communications}
\author{Jingjing Zhao, Haowen Song, Xidong Mu, Kaiquan Cai, Yanbo Zhu, and Yuanwei Liu,~\IEEEmembership{Fellow,~IEEE}
\thanks{J. Zhao, H. Song, K. Cai, Y. Zhu are with the School of Electronics and Information Engineering, Beihang University, 100191, Beijing, China, and also with the State Key Laboratory of CNS/ATM, 100191, Beijing, China. (e-mail:\{jingjingzhao, haowensong, ckq, zyb\}@buaa.edu.cn). 

X. Mu is with the Centre for Wireless Innovation (CWI), Queen's University Belfast, Belfast, BT3 9DT, U.K. (e-mail: x.mu@qub.ac.uk). 

Y. Liu is with the Department of Electrical and Electronic Engineering, the University of Hong Kong, Hong Kong, China (e-mail: yuanwei@hku.hk). }}

\maketitle
\begin{abstract}
A continuous aperture array (CAPA)-based secure communication system is investigated, where a base station (BS) equipped with a CAPA transmits signals to a legitimate user under the existence of an eavesdropper. For improving the secrecy performance, the artificial noise (AN) is employed at the BS for the jamming purpose. We aim at maximizing the secrecy rate by jointly optimizing the information-bearing and AN source current patterns, subject to the maximum transmit power constraint. To solve the resultant non-convex integral-based functional programming problem, a channel subspace-based approach is first proposed via exploiting the result that the optimal current patterns always lie within the subspace spanned by all users' channel responses. Then, the intractable CAPA continuous source current pattern design problem with an infinite number of optimization variables is equivalently transformed into the channel-subspace weighting factor optimization problem with a finite number of optimization variables. A penalty-based successive convex approximation method is developed for iteratively optimizing the finite-size weighting vectors. To further reduce the computational complexity, we propose a two-stage source current patterns design scheme. Specifically, the information-bearing and AN patterns are first designed using the maximal ration transmission (MRT) and zero-forcing (ZF) transmission, respectively. Then, the remaining power allocation is addressed via the one-dimensional search method. Numerical results unveil that 1) the CAPA brings in significant secrecy rate gain compared to the conventional discrete multiple-input multiple-output (MIMO); 2) the proposed channel subspace-based algorithm outperforms the conventional Fourier-based approach, while sustaining much lower computational complexity; and 3) the two-stage ZF-MRT approach has negligible performance loss for the large transmit power regime.
\end{abstract}

\begin{IEEEkeywords}
Continuous aperture array (CAPA), beamforming optimization, physical layer security,  artificial noise.
\end{IEEEkeywords}

\vspace{-0.2cm}
\section{Introduction}
Among wireless communication technologies, multiple-input multiple-output (MIMO) is the key enabling technology for boosting the wireless network capacity~\cite{6824752}. 
Its core principle relies on leveraging a number of antennas to effectively enhance the spatial degrees of freedom (DoFs) and provide substantial array gains. However, the performance of the conventional MIMO with spatially discrete antennas is inherently limited by the antenna density and the aperture size. As a remedy, high-density antenna architectures have emerged, such as holographic MIMO~\cite{9136592,9716880}, large intelligent surfaces (LIS)~\cite{8319526,9139337}, and dynamic metasurface antennas~\cite{9324910}. As an ultimate goal of these array architectures, the continuous aperture array (CAPA) enables (nearly) continuous electromagnetic (EM) aperture coverage with a (virtually) infinite number of radiating elements coupled with electronic circuits~\cite{liu2024capa}. It can provide precise control of the amplitude and phase of the source current across the continuous surface, thereby significantly enhancing spatial DoFs. Moreover, unlike conventional discrete MIMO where each antenna requires dedicated hardware (i.e., radio frequency (RF) ports or phase shifters), the CAPA is more cost-effective as it only requires the same number of RF ports to that of multiplexed signals~\cite{bjornson2019massive}. Therefore, the CAPA has been regarded as a revolutionary technology for future wireless networks.

Moreover, due to the broadcast nature of wireless channels, the physical layer security (PLS) is a critical issue and has attracted significant research attention~\cite{chen2016survey}. Thanks to the capability of the MIMO technique for steering signals toward desired directions, the secrecy performance can be enhanced with the effective beamforming~\cite{wyner1975wire}. With the enhanced DoFs, the CAPA is expected to provide new opportunities for the secure communications design. By properly designing the source current patterns, the CAPA can effectively boost the communication quality for legitimate information
receivers (IRs), while weakening the wiretapping quality at the eavesdroppers (Eves), and thus offer enhanced security against eavesdropping threats.

\subsection{Prior Works}
\textit {1) Studies on CAPA-based Communications:} 
In recent years, growing research interests have been devoted into CAPA-based communications. For example, the authors of~\cite{29386} demonstrated that the DoFs were practically equivalent to the Nyquist number, which corresponded to the effective bandwidth of the scattered EM field and the extension of the observation domain. From the EM formulation, the authors of~\cite{miller2000communicating} calculated the communication DoFs between two arbitrarily shaped and positioned volumes by solving eigenfunction problems, demonstrating that communication DoFs were proportional to the volumes of the transmit and receive CAPAs. In a separate study, a spatial bandwidth-based approach was developed in~\cite{9848802}, providing closed-form approximations for achievable spatial DoFs between two CAPAs in line-of-sight-dominant channels. For the analysis of the channel capacity in CAPA systems, the authors of~\cite{4685903} derived the Shannon information capacity in spacetime
wireless channels based on Maxwell’s equations. In~\cite{4447351}, the eigenfunction method was proposed to represent the capacity between two CAPAs, revealing the bounded capacity with limited power. Inspired by the Kolmogorov information theory, the authors of~\cite{8585146} characterized the channel capacity between two CAPAs to quantify the maximum information that could be reliably transmitted. Moreover, the authors of~\cite{10050124} proposed a rigorous analytical scheme for the channel capacity between two CAPAs in both random fields and non-white noise fields. As a further advancement, the authors of~\cite{ouyang2024diversity} proposed a general fading model applied to multipath channels between two CAPAs, and analyzed the diversity and multiplexing performance over fading channels. The authors of~\cite{zhao2024continuous} investigated the uplink multi-user capacity of the CAPA system and analyzed the downlink multi-user capacity by leveraging the principle of uplink-downlink duality. 

Besides the performance analysis, the beamforming design in CAPA-based communications has been studied in~\cite{9641865,9906802,10158997,liu2024holographic,10612761}. In~\cite{9641865}, the authors employed a series of square-wave functions to generate the source current patterns, showing that the quasi-optimal communication performance could be realized. The authors of~\cite{9906802} introduced a wavenumber-division multiplexing scheme to directly generate transmit current patterns and received fields by Fourier basis functions. Furthermore, the authors of~\cite{10158997} proposed a Fourier-based discretization approach for the CAPA beamforming design, with the aim of maximizing the weighted sum rate. The authors of~\cite{liu2024holographic} proposed a continuous-discrete transformation based on the Fourier transform for the integrated sensing and communication system, converting the continuous pattern design into an equivalent discrete beamforming design. Moreover, the authors of~\cite{10612761} proposed an iterative water-filling algorithm to maximize the spectral efficiency of the CAPA system, which transformed the optimization variables from the continuous source current patterns into expansion coefficients using Fourier space series. The authors of~\cite{10910020} further employed the calculus of variation theory to directly optimize the continuous source current patterns for the maximization of the weighted sum rate of CAPA-based multi-user communication systems.

\textit {2) Studies on Beamforming-based PLS:} In recent years, multiple-antenna technologies have been recognized as a promising solution for enhancing the PLS in wireless communications~\cite{7081071}. In~\cite{6584932}, the alternating optimization approach was proposed for maximizing the secrecy capacity of an MIMO channel overheard by one or multiple Eves. The secrecy capacity was characterized in terms of generalized eigenvalues in~\cite{5485016}, showing that the beamforming strategy could effectively realize the optimal capacity. As a further advance, the authors of~\cite{9733424} utilized the additional DoFs provided by reconfigurable intelligent surfaces to further enhance the secrecy performance of communication systems. Artificial noise is also widely used in PLS related researches to deteriorate the quality of the received signals at Eves and improve the secrecy rate~\cite{8543573}. In such strategies, in order to avoid interfering with IRs, a simple method is to let the AN lie in the null space of the signal space~\cite{6826572}.  However, this method often fails to achieve the optimal secrecy performance. Therefore, the authors of~\cite{6482662} applied the semidefinite programming (SDP) method to jointly optimize the beamforming weights of the information and the AN signals, with the objective of maximizing the secrecy rate. The authors of~\cite{zhao2024physical} was the first to analyze the secrecy performance achieved by CAPAs, where the closed-form expressions for the maximum secrecy rate and minimum required power were derived. 

\subsection{Motivations and Contributions}
Despite the aforementioned studies on CAPA-based communications, the research on the beamforming design for enhancing the PLS in CAPA systems is still in its infancy. On the one hand, the benefits brought by the AN-assisted secure beamforming strategy in CAPA systems remain unclear. On the other hand, due to the spatially continuous source current patterns distributed over the aperture, the CAPA secure beamforming optimization becomes more challenging compared to the conventional discrete matrices/vectors optimization. Although the Fourier-based approach is a commonly used approach for addressing the integral-based functional programming problem, it requires the approximation of the original continuous current patterns with a number of Fourier basis functions~\cite{10158997}. If the number of Fourier basis functions is chosen to be sufficiently large, the computational complexity will become prohibitively high. On the contrary, if the number of Fourier basis functions is restricted, the accurate approximation can not be derived . 

To address the above issues, we propose a novel channel-subspace based approach for the CAPA-based secure beamforming design, based on the proof that the optimal current patterns lie within the subspace spanned by the channel responses. By converting the optimization variables from continuous source current patterns to discrete channel subspace weighting factors, the conventional discrete-vectors optimization methods successive convex approximation (SCA) is employed for iteratively searching for the suboptimal solutions. Compared to the Fourier-based approach~\cite{10158997}, the proposed channel subspace-based approach effectively avoids the approximation process, and therefore can improve the optimization accuracy and reduce the computational complexity. We further propose a low-complexity two-stage beamforming algorithm based on the zero-forcing (ZF) and the maximum ratio transmission (MRT) schemes, where the source current patterns can be derived in the closed form and then only the power allocation between the two signals is optimized. In summary, the main contributions of this paper can be summarized as follows:
\begin{itemize}
\item We investigate the CAPA-based secure communications scheme, where the base station (BS) equipped with a CAPA transmits information signals toward the IR and AN signals toward the Eve for the jamming purpose. Based on the EM theory, we formulate the secrecy rate maximization problem by optimizing the source current patterns of both the information-bearing and AN signals. 

\item We propose a novel channel subspace-based method for solving the resultant non-convex integral-based functional programming optimization problem. Specifically, by exploiting the subspace spanned by all users’ channel responses, the continuous source current patterns optimization problem is first equivalently converted to the channel-subspace weighting factors optimization problem. Subsequently, the successive convex approximation (SCA) method is invoked for tackling the vector-based optimization problem. 

\item To further reduce the computational complexity, we propose a two-stage source current patterns design scheme. Specifically, the interference generated by the AN at the IR is fully eliminated with the zero-forcing (ZF) transmission, while the strength of the information signal is maximized at the IR with the maximum-ratio transmission (MRT). Subsequently, the one-dimensional search method is applied to address the remaining power allocation problem.

\item Numerical results unveil that 1) the CAPA brings in significant secrecy rate gain compared to the discrete MIMO; 2) the proposed channel subspace-based beamforming method achieves performance improvements compared to the conventional Fourier-based approach, while sustaining much lower computational complexity; and 3) the performance of the ZF-MRT approach gradually approaches that of the Fourier-based approach with the increment of the SNR.
 
\end{itemize}

\subsection{Organization and Notation}
The rest of this paper is organized as follows. Secion II introduces the CAPA-based secure communications system model and formulates the secrecy rate maximization problem. Section III proposes the channel subspace-based beamforming scheme, which is followed by the low-complexity two-stage ZF-MRT algorithm design in Section IV. Section V presents simulation results, and Section VI concludes the paper.

\textit{Notations:} Scalars, vectors and matrices are represented by regular, boldface lower case, and upper case letters, respectively. Surfaces are represented by calligraphic letters. The set of complex and real numbers are denoted by $\mathbb{C}$ and $\mathbb{R}$, respectively. The inverse, conjugate,
transpose, conjugate transpose, trace and rank operators are denoted by $(\cdot)^{-1}$, $(\cdot)^{*}$, $(\cdot)^{T}$, $(\cdot)^{H}$, $\operatorname{Tr}(\cdot)$ and $\mathrm{Rank}\left(\cdot\right)$, respectively. The Lebesgue measure of a surface $\mathcal{S}$ is denoted by $\left|\mathcal{S}\right|$. The absolute value, Euclidean norm, nuclear norm and spectral norm are denoted by $\left|\cdot\right|$, $\left\|\cdot\right\|$, $\left\|\cdot\right\|_*$ and $\left\|\cdot\right\|_2$, respectively. $\mathbf{A}\succeq0$ indicates that $\mathbf{A}$ is a positive semidefinite matrix. The ceiling operator is denoted by $\left\lceil\cdot\right\rceil$. An identity matrix of dimension $N\times N$ is denoted by $\mathbf{I}_{N}$. The big-O notations is denoted by $O(\cdot)$.

\section{System Model and Problem Formulation}
In this section, we present the CAPA-based secure communication system model, and formulate the source current patterns optimization problem for maximizing the secrecy rate.
\subsection{System Model}
\begin{figure}
	\centering
	\includegraphics[scale=0.8]{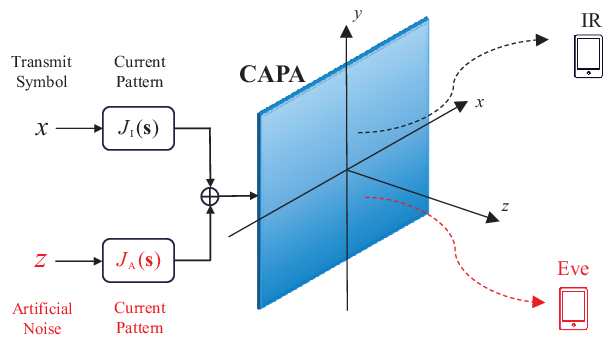}
	\caption{Illustration of the CAPA-based secure communication system.}
	\label{fig:systemmodel} 
\end{figure}

As illustrated in Fig.~\ref{fig:systemmodel}, the considered CAPA-based secure communication system consists of a BS equipped with a CAPA, a single-antenna legitimate IR, and a single-antenna Eve. Without loss of generality, we assume that the CAPA is deployed in the $x-y$ plane and centered at the origin of the coordinate system. The CAPA is with a continuous surface $\mathcal{S}_{\mathrm{T}}$ with an area of $A_{\mathrm{T}}=|\mathcal{S}_{\mathrm{T}}|$, which contains sinusoidal source currents to emit EM waves for wireless communication. Let $\mathbf{J}(\mathbf{s},\omega)\in\mathbb{C}^{3\times1}$ denote the Fourier transform of the source current density at point $\mathbf{s}\in\mathcal{S}_{\mathrm{T}}$, where $\omega=\frac{2\pi f}{c}=\frac{2\pi}{\lambda}$ denotes the angular frequency, $f $ is the signal frequency, and $\lambda $ is the signal wavelength. In this work, we focus on a narrowband single-carrier communication system where the explicit dependency of the source current on $\omega $ can be omitted. As such, the source current expression is simplified as $\mathbf{J}(\mathbf{s})$, which can be decomposed into orthogonal components along the $x $-, $y $-, and $x $-axes as follows:
\begin{align}
\mathbf{J}\left(\mathbf{s}\right)=J_x\left(\mathbf{s}\right)\hat{\mathbf{u}}_x+J_y\left(\mathbf{s}\right)\hat{\mathbf{u}}_y+J_z\left(\mathbf{s}\right)\hat{\mathbf{u}}_z,
\end{align}
where $\hat{\mathbf{u}}_x\in\mathbb{R}^{3\times1}$, $\hat{\mathbf{u}}_y\in\mathbb{R}^{3\times1}$, and $\hat{\mathbf{u}}_z\in\mathbb{R}^{3\times1}$ are unit vectors along the $x $-, $y $-, and $x $-axes, respectively. Here, we consider the case of a vertically polarized transmitter, where only the $y $-component of the source current is excited. Therefore, the source current can be further simplified as
\begin{align}
\label{eq:source current1}
\mathbf{J}\left(\mathbf{s}\right)=J\left(\mathbf{s}\right)\hat{\mathbf{u}}_y,
\end{align}
where we define the scalar source current $J\left(\mathbf{s}\right) := J_y(\mathbf{s})$ to simplify the notation. Specifically, let $x\in\mathbb{C}$ and $z\in\mathbb{C}$ denote the information and the AN signals, respectively. Note that $\mathbb{E}\{\left|x\right|^{2}\}=1$ and $\mathbb{E}\{\left|z\right|^{2}\}=1$. Then, $J\left(\mathbf{s}\right)$ can be represented  by the linear superposition of the information-bearing source current $J_\mathrm{I}\left(\mathbf{s}\right)\in\mathbb{C}$ and the AN source current $J_\mathrm{A}\left(\mathbf{s}\right)\in\mathbb{C}$, represented by
\begin{align}
\label{eq:source current2}
J\left(\mathbf{s}\right)=J_\mathrm{I}\left(\mathbf{s}\right)x+J_\mathrm{A}\left(\mathbf{s}\right)z.
\end{align}

Let $\mathbf{r}_\text{I}\in\mathbb{C}^{3\times1}$ and $\mathbf{r}_\text{E}\in\mathbb{C}^{3\times1}$ denote the positions of the IR and the Eve, respectively. By introducing the
Green’s function, the electric fields generated at the positions $\mathbf{r}_\mathrm{I}$ and $\mathbf{r}_\mathrm{E}$ by the source current $\mathbf{J}\left(\mathbf{s}\right)$ in a homogeneous medium are respectively given by~\cite{9139337},~\cite{poon2005degrees}:
\begin{subequations}
\begin{equation}
\mathbf{E}_\mathrm{I}=\int_{\mathcal{S}_\mathrm{T}}\mathbf{G}\left(\mathbf{r}_\mathrm{I},\mathbf{s}\right)\mathbf{J}\left(\mathbf{s}\right)d\mathbf{s}\in\mathbb{C}^{3\times1},
\end{equation}
\begin{equation}
\mathbf{E}_\mathrm{E}=\int_{\mathcal{S}_\mathrm{T}}\mathbf{G}\left(\mathbf{r}_\mathrm{E},\mathbf{s}\right)\mathbf{J}\left(\mathbf{s}\right)d\mathbf{s}\in\mathbb{C}^{3\times1}.
\end{equation}
\end{subequations}
From the perspective of mathematics, the integral kernel $\mathbf{G}\left(\mathbf{r},\mathbf{s}\right)$ is typically referred to as the Green's function. In ideal
unbounded and homogeneous mediums, $\mathbf{G}\left(\mathbf{r},\mathbf{s}\right)$ can be expressed as
\begin{align}
\mathbf{G}\left(\mathbf{r},\mathbf{s}\right)=-\frac{j\eta e^{-j\frac{2\pi}{\lambda}\|\mathbf{r}-\mathbf{s}\|}}{2\lambda\|\mathbf{r}-\mathbf{s}\|}\left(\mathbf{I}_3-\frac{(\mathbf{r}-\mathbf{s})(\mathbf{r}-\mathbf{s})^T}{\|\mathbf{r}-\mathbf{s}\|^2}\right),
\end{align}
where $\eta $ represents the intrinsic impedance. 

We consider that practical single-polarized antennas are equipped at both the IR and the Eve, with polarization directions $\hat{\mathbf{u}}_\mathrm{I}\in\mathbb{R}^{3\times1}$ and $\hat{\mathbf{u}}_\mathrm{E}\in\mathbb{R}^{3\times1}$, respectively, satisfying $\left\|\hat{\mathbf{u}}_\mathrm{I}\right\|=\left\|\hat{\mathbf{u}}_\mathrm{E}\right\|=1$. As such, the electric field captured by the IR and the Eve are respectively given by
\begin{align}
\label{eq:field1}
E_\mathrm{I}=\hat{\mathbf{u}}_\mathrm{I}^T\mathbf{E}_\mathrm{I}+n_\mathrm{I}=\int_{\mathcal{S}_\mathrm{T}}\hat{\mathbf{u}}_\mathrm{I}^T\mathbf{G}\left(\mathbf{r}_\mathrm{I},\mathbf{s}\right)\mathbf{J}\left(\mathbf{s}\right)d\mathbf{s}+n_\mathrm{I},
\end{align}
and
\begin{align}
\label{eq:field2}
E_\mathrm{E}=\hat{\mathbf{u}}_\mathrm{E}^T\mathbf{E}_\mathrm{E}+n_\mathrm{E}=\int_{\mathcal{S}_\mathrm{T}}\hat{\mathbf{u}}_\mathrm{E}^T\mathbf{G}\left(\mathbf{r}_\mathrm{E},\mathbf{s}\right)\mathbf{J}\left(\mathbf{s}\right)d\mathbf{s}+n_\mathrm{E},
\end{align}
where $n_\mathrm{I}\in\mathbb{C}$ and $n_\mathrm{E}\in\mathbb{C}$ represent the EM noise at the IR and the Eve, respectively,  which can be modeled as independent Gaussian variables with zero mean and variance $\sigma_\mathrm{I}^2$ and $\sigma_\mathrm{E}^2$, i.e., $n_\mathrm{I}\sim\mathcal{CN}\left(0,\sigma_\mathrm{I}^2\right)$ and $n_\mathrm{E}\sim\mathcal{CN}(0,\sigma_\mathrm{E}^2)$~\cite{10158997}. Substituting~\eqref{eq:source current1} into~\eqref{eq:field1} and ~\eqref{eq:source current2} into ~\eqref{eq:field2}, respectively, yields
\begin{align}
E_\mathrm{I}=\int_{\mathcal{S}_\mathrm{T}}H_\mathrm{I}\left(\mathbf{s}\right)J_\mathrm{I}\left(\mathbf{s}\right)xd\mathbf{s}+\int_{\mathcal{S}_\mathrm{T}}H_\mathrm{I}\left(\mathbf{s}\right)J_\mathrm{A}\left(\mathbf{s}\right)zd\mathbf{s}+n_{I},
\end{align}
\begin{align}
E_\mathrm{E}=\int_{\mathcal{S}_\mathrm{T}}H_\mathrm{E}\left(\mathbf{s}\right)J_\mathrm{I}\left(\mathbf{s}\right)xd\mathbf{s}+\int_{\mathcal{S}_\mathrm{T}}H_\mathrm{E}\left(\mathbf{s}\right)J_\mathrm{A}\left(\mathbf{s}\right)zd\mathbf{s}+n_{E},
\end{align}
where $H_\mathrm{I}\left(\mathbf{s}\right)=\hat{\mathbf{u}}_\mathrm{I}^T\mathbf{G}\left(\mathbf{r}_\mathrm{I},\mathbf{s}\right)\hat{\mathbf{u}}_y$ and $H_\mathrm{E}\left(\mathbf{s}\right)=\hat{\mathbf{u}}_\mathrm{E}^T\mathbf{G}\left(\mathbf{r}_\mathrm{E},\mathbf{s}\right)\hat{\mathbf{u}}_y$ represent the continuous EM channels for the IR and the Eve, respectively.

The signal-to-interference-plus-noise ratio (SINR) for decoding the desired signal at the IR and Eve are respectively given by
\begin{align}
\gamma_\mathrm{I}=\frac{\left|\int_{\mathcal{S}_\mathrm{T}}H_\mathrm{I}^*\left(\mathbf{s}\right)J_\mathrm{I}\left(\mathbf{s}\right)d\mathbf{s}\right|^2}{\left|\int_{\mathcal{S}_\mathrm{T}}H_\mathrm{I}^*\left(\mathbf{s}\right)J_\mathrm{A}\left(\mathbf{s}\right)d\mathbf{s}\right|^2+\sigma_\mathrm{I}^2},
\end{align}
\begin{align}
\gamma_\mathrm{E}=\frac{\left|\int_{\mathcal{S}_\mathrm{T}}H_\mathrm{E}^*\left(\mathbf{s}\right)J_\mathrm{I}\left(\mathbf{s}\right)d\mathbf{s}\right|^2}{\left|\int_{\mathcal{S}_\mathrm{T}}H_\mathrm{E}^*\left(\mathbf{s}\right)J_\mathrm{A}\left(\mathbf{s}\right)d\mathbf{s}\right|^2+\sigma_\mathrm{E}^2},
\end{align}
The achievable rates are thus given by $R_\mathrm{I}=\operatorname{log}_2\left(1+\gamma_\mathrm{I}\right)$ and $R_\mathrm{E}=\operatorname{log}_2\left(1+\gamma_\mathrm{E}\right)$. Then, the secrecy
rate is given by
\begin{align}
\label{eq:RS}
R_\mathrm{S}=\left[R_\mathrm{I}-R_\mathrm{E}\right]^+.
\end{align}

\subsection{Problem Formulation}
In this paper, we aim to maximize the secrecy rate by jointly optimizing the source current patterns $J_\mathrm{I}(\mathbf{s})$ and $J_\mathrm{A}(\mathbf{s})$, subject to the BS maximum transmit power constraint $P_\mathrm{T}$. The optimization problem is formulated as
\begin{subequations}
\label{eq:optimization_problem}
\begin{equation}
\label{eq:15aSRM_objective}
\max_{J_\mathrm{I}(\mathbf{s}),J_\mathrm{A}(\mathbf{s})}
R_\mathrm{S}
\end{equation}
\begin{equation}
\label{eq:15power_constraint}
    {\rm{s.t.}} \ \ \int_{\mathcal{S}_\mathrm{T}}\left|J_\mathrm{I}\left(\mathbf{s}\right)\right|^2+\left|J_\mathrm{A}\left(\mathbf{s}\right)\right|^2d\mathbf{s}\leq P_\mathrm{T}.
\end{equation}
\end{subequations}
Notably, Problem~\eqref{eq:optimization_problem} is a non-convex integral-based functional programming problem, which is intractable to address compared to the conventional discrete MIMO beamforming problems due to the following two reasons. First, the optimization variables $J_\mathrm{I}\left(\mathbf{s}\right)$ and $J_\mathrm{A}\left(\mathbf{s}\right)$ are continuous with infinite dimensions. Second, both the objective function and the constraint are with integral forms, which results in high optimization complexity mathematically. To address the above challenges, we propose a novel channel subspace-based approach and a low-complexity two-stage ZF-MRT approach in the following. 

\section{Channel Subspace-Based Optimization Approach}
In this section, by representing the optimal source current patterns $J_\mathrm{I}\left(\mathbf{s}\right)$ and $J_\mathrm{A}\left(\mathbf{s}\right)$ as linear combinations of the continuous EM channels of the IR and the Eve, we transform the original integral-based functional programming problem to a discrete vector-based optimization problem. 

\subsection{Problem Reformulation Based on Channel Subspace}
We first employ the following lemma to express the optimal $J_\mathrm{I}\left(\mathbf{s}\right)$ and $J_\mathrm{A}\left(\mathbf{s}\right)$ as linear combinations of $H_\mathrm{I}\left(\mathbf{s}\right)$ and $H_\mathrm{E}\left(\mathbf{s}\right)$.

\textbf{Lemma 1.} The optimal $J_\mathrm{I}\left(\mathbf{s}\right)$ and $J_\mathrm{A}\left(\mathbf{s}\right)$ can be expressed as
\begin{align}
\label{eq:linerJI}
J_\mathrm{I}\left(\mathbf{s}\right)=\alpha_{11}H_\mathrm{I}\left(\mathbf{s}\right) + \alpha_{12}H_\mathrm{E}\left(\mathbf{s}\right),
\end{align}
\begin{align}
\label{eq:linerJA}
J_\mathrm{A}(\mathbf{s})=\alpha_{21}H_\mathrm{I}(\mathbf{s}) + \alpha_{22}H_\mathrm{E}(\mathbf{s}),
\end{align}
where ${\alpha_{11}, \alpha_{12}, \alpha_{21}, \alpha_{22}}$ are weighting coefficients.

\begin{proof}
    Please refer to Appendix A.
\end{proof}

\textbf{Lemma 1} indicates that once $H_{I}\left(\mathbf{s}\right)$ and $H_{E}\left(\mathbf{s}\right)$ are known, the optimal source current patters $J_\mathrm{I}\left(\mathbf{s}\right)$ and $J_\mathrm{A}\left(\mathbf{s}\right)$ can be obtained by optimizing the following weighting vectors:
\begin{align}
\label{eq:alpha}
& \mathbf{a}_{1}=[\alpha_{11},\alpha_{12}]^{T}, \mathbf{a}_{2}=[\alpha_{21},\alpha_{22}]^{T}.
\end{align}
To this end, the optimization variables can be transformed from continuous functions $J_\mathrm{I}\left(\mathbf{s}\right)$ and $J_\mathrm{A}\left(\mathbf{s}\right)$ into finite-dimensional vectors $\mathbf{a}_{1}$ and $\mathbf{a}_{2}$. 
Specifically, define the channel correlation matrix $\mathbf{H}_{\text{cor}}$ as
\begin{align}
\label{eq:CR}
\mathbf{H}_{\text{cor}} =
\begin{bmatrix}
h_\mathrm{I,I} & h_\mathrm{I,E} \\
h_\mathrm{E,I} & h_\mathrm{E,E}
\end{bmatrix},
\end{align}
where $h_{i,j}$ is given by
 \begin{align}
 \label{eq:h}
h_{i,j}\triangleq\int_{\mathcal{S}_{\mathrm{T}}}H_j\left(\mathbf{s}\right)H_i^*\left(\mathbf{s}\right)d\mathbf{s},\forall i,j.
\end{align}
Substituting~\eqref{eq:linerJI} and~\eqref{eq:linerJA} into $R_\mathrm{I}=\operatorname{log}_2(1+\gamma_\mathrm{I})$ yields
\begin{align}
\label{eq:RI}
& R_\mathrm{I} = \nonumber\\
& \operatorname{log}_2\left(1+\frac{\left|\int_{\mathcal{S}_\mathrm{T}}H_\mathrm{I}^*\left(\mathbf{s}\right)\left(\alpha_{11}H_\mathrm{I}\left(\mathbf{s}\right) + \alpha_{12}H_\mathrm{E}\left(\mathbf{s}\right)\right)d\mathbf{s}\right|^2}{\left|\int_{\mathcal{S}_\mathrm{T}}H_\mathrm{I}^*\left(\mathbf{s}\right)\left(\alpha_{21}H_\mathrm{I}\left(\mathbf{s}\right) + \alpha_{22}H_\mathrm{E}\left(\mathbf{s}\right)\right)d\mathbf{s}\right|^2 + \sigma_\mathrm{I}^2}\right)\nonumber\\
& \overset{(a)}{\operatorname*{\operatorname*{=}}}\operatorname{log}_2\left(1+\frac{\left|\alpha_{11} h_\mathrm{I,I}^* + \alpha_{12} h_\mathrm{E,I}^*\right|^2}{\left|\alpha_{21}h_\mathrm{I,I}^* + \alpha_{22}h_\mathrm{E,I}^*\right|^2 + \sigma_\mathrm{I}^2}\right)\nonumber\\
& = \operatorname{log}_2\left(1+\frac{\left|\mathbf{h}_\mathrm{I}^{H}\mathbf{a}_{1}\right|^{2}}{\left|\mathbf{h}_\mathrm{I}^{H}\mathbf{a}_{2}\right|^{2} + \sigma_\mathrm{I}^2}\right),
\end{align}
where 
\begin{align}
& \mathbf{h}_\mathrm{I}=\left[h_\mathrm{I,I},h_\mathrm{E,I}\right]^{T}, \mathbf{h}_\mathrm{E}=\left[h_\mathrm{I,E},h_\mathrm{E,E}\right]^{T}.
\end{align}
Equation $(a)$ is due to the fact that $h_{i,j}=h_{j,i}^*$. Similarly, $R_\mathrm{E}$ formed by the optimal $J_\mathrm{I}\left(\mathbf{s}\right)$ and $J_\mathrm{A}\left(\mathbf{s}\right)$ is given by
\begin{align}
\label{eq:RE}
R_\mathrm{E} = \operatorname{log}_2\left(1+\frac{\left|\mathbf{h}_\mathrm{E}^{H}\mathbf{a}_{1}\right|^{2}}{\left|\mathbf{h}_\mathrm{E}^{H}\mathbf{a}_{2}\right|^{2} + \sigma_\mathrm{E}^2}\right).
\end{align}

Accordingly, the original optimization problem~\eqref{eq:optimization_problem} can be transformed to the following equivalent form
\begin{subequations}
\label{eq:optimization_problem_discrete}
\begin{equation}
\label{eq:24aSRM_objective}
\max_{\{\mathbf{a}_i\}_{i=1,2}}
f\left(\mathbf{a}_{1},\mathbf{a}_{2}\right)
\end{equation}
\begin{equation}
\label{eq:24power_constraint}
{\rm{s.t.}} \ \ \mathbf{a}_{1}^H\mathbf{H}_{\text{cor}}\mathbf{a}_{1}+\mathbf{a}_{2}^H\mathbf{H}_{\text{cor}}\mathbf{a}_{2}\leq P_\mathrm{T},
\end{equation}
\end{subequations}
where $f\left(\mathbf{a}_{1},\mathbf{a}_{2}\right)$ is defined as~\eqref{eq:f} at the top of the next page. Although the terms of integrals have been explicitly removed in problem~\eqref{eq:optimization_problem_discrete}, it is still challenging to solve due to the non-convexity with respect to $\mathbf{a}_{1}$ and $\mathbf{a}_{2}$. In the following, we propose a penalty-based SCA approach for iteratively finding the sub-optimal solution. 
\begin{figure*}
\begin{equation}
\label{eq:f}
f\left(\mathbf{a}_{1},\mathbf{a}_{2}\right)=
\operatorname{log}_2\left(1+\frac{\left|\mathbf{h}_\mathrm{I}^{H}\mathbf{a}_{1}\right|^{2}}{\left|\mathbf{h}_\mathrm{I}^{H}\mathbf{a}_{2}\right|^{2} + \sigma_\mathrm{I}^2}\right)-\operatorname{log}_2\left(1+\frac{\left|\mathbf{h}_\mathrm{E}^{H}\mathbf{a}_{1}\right|^{2}}{\left|\mathbf{h}_\mathrm{E}^{H}\mathbf{a}_{2}\right|^{2} + \sigma_\mathrm{E}^2}\right).
\end{equation}
\end{figure*}

\subsection{Penalty-Based SCA Algorithm}
To overcome the quadratic forms, we set $\mathbf{H}_\mathrm{I} = \mathbf{h}_\mathrm{I}\mathbf{h}_\mathrm{I}^H$,  $\mathbf{H}_\mathrm{E} = \mathbf{h}_\mathrm{E}\mathbf{h}_\mathrm{E}^H$, and $\mathbf{A}_i = \mathbf{a}_i\mathbf{a}_i^H$, $i=1, 2$, which satisfies $\mathbf{A}_i\succeq0$ and $\mathrm{Rank}\left(\mathbf{A}_i\right)=1$. Substituting~\eqref{eq:RI} and~\eqref{eq:RE}  into~\eqref{eq:RS}, $R_\mathrm{S}$ can be rewritten as given in~\eqref{eq:29} at the top of the next page,
\begin{figure*}[!h]
    \begin{align}
     R_\mathrm{S}&= \operatorname{log}_2\left(1+\frac{\left|\mathbf{h}_\mathrm{I}^{H}\mathbf{a}_{1}\right|^{2}}{\left|\mathbf{h}_\mathrm{I}^{H}\mathbf{a}_{2}\right|^{2} + \sigma_\mathrm{I}^2}\right) - \operatorname{log}_2\left(1+\frac{\left|\mathbf{h}_\mathrm{E}^{H}\mathbf{a}_{1}\right|^{2}}{\left|\mathbf{h}_\mathrm{E}^{H}\mathbf{a}_{2}\right|^{2} + \sigma_\mathrm{E}^2}\right)\nonumber\\
& \overset{(b)}{\operatorname*{\operatorname*{=}}} \operatorname{log}_2\left(\frac{\mathrm{Tr}\left(\mathbf{H}_\mathrm{I}\mathbf{A}_1\right) + \mathrm{Tr}\left(\mathbf{H}_\mathrm{I}\mathbf{A}_2\right) + \sigma_\mathrm{I}^2}{\mathrm{Tr}\left(\mathbf{H}_\mathrm{I}\mathbf{A}_2\right) + \sigma_\mathrm{I}^2}\right)
 - \operatorname{log}_2\left(\frac{\mathrm{Tr}\left(\mathbf{H}_\mathrm{E}\mathbf{A}_1\right) + \mathrm{Tr}\left(\mathbf{H}_\mathrm{E}\mathbf{A}_2\right)+ \sigma_\mathrm{E}^2}{\mathrm{Tr}\left(\mathbf{H}_\mathrm{E}\mathbf{A}_2\right) + \sigma_\mathrm{E}^2}\right).
    \label{eq:29}
\end{align}
\hrulefill
\end{figure*}
where equality $(b)$ holds because $\mathrm{Tr}\left(\mathbf{AB}\right) = \mathrm{Tr}\left(\mathbf{BA}\right)$.
To facilitate the design, we introduce the exponential auxiliary variables $\tau,\varepsilon,u,v$ as follows:
\begin{align}
e^{\tau}=\mathrm{Tr}\left(\mathbf{H}_\mathrm{I}\mathbf{A}_1\right) + \mathrm{Tr}\left(\mathbf{H}_\mathrm{I}\mathbf{A}_2\right) + \sigma_\mathrm{I}^2,
\end{align}
\begin{align}
\label{eq:auxiliary_variables_ep}
e^{\varepsilon}=\mathrm{Tr}\left(\mathbf{H}_\mathrm{I}\mathbf{A}_2\right) + \sigma_\mathrm{I}^2,
\end{align}
\begin{align}
\label{eq:auxiliary_variables_u}
e^{u}=\mathrm{Tr}\left(\mathbf{H}_\mathrm{E}\mathbf{A}_1\right) + \mathrm{Tr}\left(\mathbf{H}_\mathrm{E}\mathbf{A}_2\right)+ \sigma_\mathrm{E}^2,
\end{align}
\begin{align}
e^{v}=\mathrm{Tr}\left(\mathbf{H}_\mathrm{E}\mathbf{A}_2\right) + \sigma_\mathrm{E}^2.
\end{align}
Then, problem~\eqref{eq:optimization_problem} can be equivalently transformed as
\begin{subequations}
\label{eq:optimization_problem4}
\begin{equation}
\label{eq:29aSRM_objective}
\max_{\tau,\varepsilon,u,v,\{\mathbf{A}_i\}_{i=1,2}}
\tau-\varepsilon-u+v
\end{equation}
\begin{equation}
\label{eq:convex1}
{\rm{s.t.}} \ \ 
e^{\tau} \leq \mathrm{Tr}\left(\mathbf{H}_\mathrm{I}\mathbf{A}_1\right) + \mathrm{Tr}\left(\mathbf{H}_\mathrm{I}\mathbf{A}_2\right) + \sigma_\mathrm{I}^2,
\end{equation}
\begin{equation}
\label{eq:nonconvex1}
e^{\varepsilon} \geq \mathrm{Tr}\left(\mathbf{H}_\mathrm{I}\mathbf{A}_2\right) + \sigma_\mathrm{I}^2,
\end{equation}
\begin{equation}
\label{eq:nonconvex2}
e^{u} \geq \mathrm{Tr}\left(\mathbf{H}_\mathrm{E}\mathbf{A}_1\right) + \mathrm{Tr}\left(\mathbf{H}_\mathrm{E}\mathbf{A}_2\right)+ \sigma_\mathrm{E}^2,
\end{equation}
\begin{equation}
\label{eq:convex2}
e^{v} \leq \mathrm{Tr}\left(\mathbf{H}_\mathrm{E}\mathbf{A}_2\right) + \sigma_\mathrm{E}^2.
\end{equation}
\begin{equation}
\label{eq:29power_constraint}
    \mathrm{Tr}\left(\mathbf{H}_{\text{cor}}\mathbf{A}_1\right) + \mathrm{Tr}\left(\mathbf{H}_{\text{cor}}\mathbf{A}_2\right)\leq P_\mathrm{T}.
\end{equation}
\begin{equation}
\label{eq:29semidefinite_constraint}
    \mathbf{A}_1\succeq0, \mathbf{A}_2\succeq0.
\end{equation}
\begin{equation}
\label{eq:29rank1_constraint}
    \mathrm{Rank}\left(\mathbf{A}_1\right)=1, \mathrm{Rank}\left(\mathbf{A}_2\right)=1,
\end{equation}
\end{subequations}
Obviously, the inequalities~\eqref{eq:convex1}-\eqref{eq:convex2} hold equality at the optimum point, which can be verified by the monotonicity of the objective function. However,~\eqref{eq:nonconvex1} and~\eqref{eq:nonconvex2} are still nonconvex. To convert the non-convex constraints into convex ones, Taylor expansion is adopted. Specifically, the first-order Taylor expansion of~\eqref{eq:nonconvex1} and~\eqref{eq:nonconvex2} can be written as
\begin{align}
\label{eq:convex3}
\mathrm{Tr}\left(\mathbf{H}_\mathrm{I}\mathbf{A}_2\right) + \sigma_\mathrm{I}^2\leq e^{\varepsilon^{\left(k\right)}}\left(\varepsilon-\varepsilon^{\left(k\right)}+1\right),
\end{align}
\begin{align}
\label{eq:convex4}
\mathrm{Tr}\left(\mathbf{H}_\mathrm{E}\mathbf{A}_1\right) + \mathrm{Tr}\left(\mathbf{H}_\mathrm{E}\mathbf{A}_2\right)+ \sigma_\mathrm{E}^2\leq e^{u^{\left(k\right)}}\left(u-u^{\left(k\right)}+1\right),
\end{align}
where $\varepsilon^{\left(k\right)}$ and $u^{\left(k\right)}$ are the optimal solutions at the $k$-th iteration. 
After the above operation, the optimization problem can be reformulated as
\begin{subequations}
\label{eq:optimization_problem2}
\begin{equation}
\label{eq:39aSRM_objective}
\max_{\tau,\varepsilon,u,v,\{\mathbf{A}_i\}_{i=1,2}}
\tau-\varepsilon-u+v
\end{equation}
\begin{equation}
\label{eq:39power_constraint}
    {\rm{s.t.}} \ \ 
    \eqref{eq:convex1},\eqref{eq:convex2},\eqref{eq:29power_constraint}-\eqref{eq:29rank1_constraint},\eqref{eq:convex3}-\eqref{eq:convex4}.
\end{equation}
\end{subequations}
we note that problem~\eqref{eq:optimization_problem2} is still nonconvex due to the rank-one constraints \eqref{eq:29rank1_constraint}. Next, we will employ a penalty-based method~\cite{ben1997penalty} to address the difficulty of the rank-one constraint.

The non-convex rank-one constraints \eqref{eq:29rank1_constraint} can be equivalently rewritten as the following equality constraint:
\begin{equation}
\label{eq:penalty}
\left\|\mathbf{A}_i\right\|_*-\left\|\mathbf{A}_i\right\|_2=0, i=1, 2,
\end{equation}
where $\left\|\mathbf{A}_i\right\|_*=\sum_j\sigma_j\left(\mathbf{A}_i\right)$ and $\left\|\mathbf{A}_i\right\|_2=\sigma_1\left(\mathbf{A}_i\right)$ denote the nuclear norm and spectral norm, respectively, and $\sigma_j\left(\mathbf{A}_i\right)$ is the $j$th largest singular value of matrix $\mathbf{A}_i$. We note that $\mathbf{A}_i = \mathbf{a}_i\mathbf{a}_i^H$ is a Hermitian matrix and $\mathbf{A}_i\succeq0$, so it always satisfies $\left\|\mathbf{A}_i\right\|_*-\left\|\mathbf{A}_i\right\|_2 \geq 0$, where quality holds only if $\mathbf{A}_i$ is a rank-one matrix.
Furthermore, we employ the penalty method to incorporate~\eqref{eq:penalty} as a penalty term into the objective function. As such, \eqref{eq:39aSRM_objective} can be rewritten as
\begin{equation}
\max_{\tau,\varepsilon,u,v,\{\mathbf{A}_i\}_{i=1,2}}
\tau-\varepsilon-u+v - \eta\sum_{i = 1,2}\left(\left\|\mathbf{A}_i\right\|_*-\left\|\mathbf{A}_i\right\|_2\right)
\end{equation}
where $\eta > 0$ is the penalty factor which penalizes the objective function if $\mathbf{A}_i$ is not rank-one. Therefore, the rank-one constraints \eqref{eq:29rank1_constraint} can be dropped.

Note that the penalty term is nonconvex and in the form of difference convex (DC) functions. For a given point $\mathrm{A}_{i}^{(k)}$ in the $k$th iteration of the SCA method, using the first-order Taylor expansion to construct the upper-bound surrogate function as follow:
\begin{equation}
\left\|\mathbf{A}_i\right\|_*-\left\|\mathbf{A}_i\right\|_2
\leq
\left\|\mathbf{A}_i\right\|_*-\overline{\mathbf{A}}_i^{(k)},
\end{equation}
where $\overline{\mathbf{A}}_i^{(k)} \triangleq\left\|\mathbf{A}_i^{(k)}\right\|_2
+
\mathrm{Tr}\left[\overline{\mathbf{x}}\left(\mathbf{A}_{i}^{(k)}\right)\left(\overline{\mathbf{x}}\left(\mathbf{A}_{i}^{(k)}\right)\right)^{H}\left(\mathbf{A}_{i}-\mathbf{A}_{i}^{(k)}\right)\right]$ and $\overline{\mathbf{x}}\left(\mathbf{A}_{i}^{(k)}\right)$ denotes the eigenvector w.r.t. the largest eigenvalue of $\mathbf{A}_{i}^{(k)}$. As a result, for given point $\mathbf{A}_{i}^{(k)}$, problem~\eqref{eq:optimization_problem2} is transformed into the following optimization problem:
\begin{subequations}
\label{eq:optimization_problem3}
\begin{equation}
\label{eq:71aSRM_objective}
\max_{\tau,\varepsilon,u,v,\{\mathbf{A}_i\}_{i=1,2}}
\tau-\varepsilon-u+v - \eta\sum_{i = 1,2}\left(\left\|\mathbf{A}_i\right\|_*-\overline{\mathbf{A}}_i^{\left(k\right)}\right)
\end{equation}
\begin{equation}
\label{eq:71power_constraint}
    {\rm{s.t.}} \ \     \eqref{eq:convex1},\eqref{eq:convex2},\eqref{eq:29power_constraint}-\eqref{eq:29semidefinite_constraint},\eqref{eq:convex3}-\eqref{eq:convex4}.
\end{equation}
\end{subequations}
Based on the above, it can be verified that as $\eta\to+\infty$, the solution to problem~\eqref{eq:optimization_problem3} always satisfies~\eqref{eq:penalty}, i.e., problems~\eqref{eq:optimization_problem3} and~\eqref{eq:optimization_problem2} are equivalent. It should be noted that, if the initial value of $\eta$ is set too large, the maximization of secrecy rate has almost no impact on the solution. To avoid this, we first initialize $\eta$ with a small value and then gradually increase $\eta$ to a sufficiently large value during the iteration process, ultimately obtaining rank-one matrices. 
This inspires us to use a penalty-based double-loop iterative algorithm to solve problem~\eqref{eq:optimization_problem3}. Specifically,  in the outer loop, the penalty factor is gradually increased during each iteration, i.e., $\eta=\omega\eta $, where $\omega>1$. The algorithm terminates when the penalty term $\left\|\mathbf{A}_i\right\|_*-\left\|\mathbf{A}_i\right\|_2$ falls below a threshold $\varepsilon_1$:
\begin{equation}
\label{eq:37}
\max\left\{\left\|\mathbf{A}_i\right\|_*-\left\|\mathbf{A}_i\right\|_2, i=1, 2\right\}\leq\varepsilon_1,
\end{equation}
where $\varepsilon_1$ denotes a predefined maximum violation of equality constraint~\eqref{eq:penalty}. In the inner loop, $\varepsilon, u$ and $\mathbf{A}_i$ are jointly optimized with given $\eta$. Specifically, for given $\{\varepsilon^{(k)}, u^{(k)}, \{\mathbf{A}_i^{(k)}\}_{i=1,2}\}$, the original problem~\eqref{eq:optimization_problem3} can be solved by applying CVX upon the relaxed semi-definite problem (SDP). $\{\varepsilon^{(k+1)}, u^{(k+1)}, \{\mathbf{A}_i^{(k+1)}\}_{i=1,2}\}$ are updated based on the optimal solution obtained in the $k$-th iteration. The inner loop continues until the increment of the objective function value is less than a predefined threshold $\varepsilon_2>0$ or the maximum number of inner iterations $k_{\mathrm{max}}$ is reached. The details of the developed algorithm are summarized in $\textbf{Algorithm 1}$.
\begin{algorithm}[h]
\renewcommand{\algorithmicrequire}{\textbf{Input:}}
\renewcommand{\algorithmicensure}{\textbf{Output:}}
\caption{Proposed Penalty-Based Iterative Algorithm}
\label{alg:doubleloop-optimization}
    \begin{algorithmic}[1]
    \Require $\mathbf{H}_{\text{cor}}$, $\mathbf{H}_\mathrm{I}$, $\mathbf{H}_\mathrm{E}$, $\varepsilon_1$, $\varepsilon_2$, $P_\mathrm{T}$
    \Ensure $\{\mathbf{A}_i\}_{i=1,2}$
    \State Initialize  feasible points $\{\varepsilon^{(0)}, u^{(0)}, \{\mathbf{A}_i^{(0)}\}_{i=1,2}\}$ and penalty factor $\eta$. 
    \Repeat \textbf{:\ outer loop}
        \State Set iteration index $k = 0$ for inner loop.
            \Repeat \textbf{:\ inner loop}
            \State For given points $\{\varepsilon^{(k)}, u^{(k)}, \{\mathbf{A}_i^{(k)}\}_{i=1,2}\}$, solve the relaxed problem~\eqref{eq:optimization_problem3}.
            \State Update $\{\varepsilon^{(k+1)}, u^{(k+1)}, \{\mathbf{A}_i^{(k+1)}\}_{i=1,2}\}$, and $k=k+1$.
            \Until the increase of the objective function value is below a threshold $\varepsilon_2$ or the maximum number of inner iterations $k_{\mathrm{max}}$ is reached.
            \State Update $\{\varepsilon^{(0)}, u^{(0)}, \{\mathbf{A}_i^{(0)}\}_{i=1,2}\}$ with the current solutions $\{\varepsilon^{(k)}, u^{(k)}, \{\mathbf{A}_i^{(k)}\}_{i=1,2}\}$ and $\eta=\omega\eta $.
    \Until $\max\left\{\left\|\mathbf{A}_i\right\|_*-\left\|\mathbf{A}_i\right\|_2, i=1, 2\right\}\leq\varepsilon_1$.
    \end{algorithmic}
\end{algorithm}

\subsection{Initialization Scheme}
For the initialization of the proposed channel subspace-based beamforming scheme, it is necessary to calculate the channel correlation coefficients $h_{i,j}$ as defined in~\eqref{eq:CR}. These integrals can be calculated using the Gauss-Legendre quadrature formula, which is defined as~\cite{10910020}
\begin{equation}
\label{eq:Gauss-Legendre}
\int_{-1}^1f\left(x\right)dx\approx\sum_{m=1}^M\omega_mf\left(x_m\right),
\end{equation}
where $M$ is the number of sample points, $\omega_m$ is the weight for the Gauss-Legendre quadrature, and $x_m$ is the Gauss point, i.e., the root of the Legendre
polynomial $P_{m+1}\left(x\right)$. The larger value of $M$ results in higher approximation accuracy. 
Let $L_{x}$ and $L_{y}$ denote the lengths of the CAPA along the $x-$ and $y-$ axes, respectively. According to~\eqref{eq:Gauss-Legendre}, $h_{i,j}$ can be calculated as
\begin{align}
h_{i,j}  = &\int_{\mathcal{S}_\mathrm{T}}H_j\left(\mathbf{s}\right)H_i^*\left(\mathbf{s}\right)d\mathbf{s}\nonumber\\
 = &\int_{-\frac{L_y}{2}}^{\frac{L_y}{2}}\int_{-\frac{L_x}{2}}^{\frac{L_x}{2}} H_j\left(s_x,s_y\right)H_i^*\left(s_x,s_y\right)ds_xds_y\nonumber\\
 \approx & \frac{L_xL_y} {4}\sum_{m_y=1}^M\sum_{m_x=1}^M\omega_{m_x}\omega_{m_y}H_j\left(\frac{x_{m_x}L_x}{2},\frac{x_{m_y}L_y}{2}\right)\nonumber\\
 & \times H_i^*\left(\frac{x_{m_x}L_x}{2},\frac{x_{m_y}L_y}{2}\right),
\end{align}
where the last step is obtained using the Gauss-Legendre quadrature. For the initialization of the weighting vectors $\mathbf{a}_{1}$ and $\mathbf{a}_{2}$ , we will discuss in details in Section V.  

\subsection{Convergence and Complexity Analysis}
We now present the convergence analysis of the proposed \textbf{Algorithm 1}. The algorithm terminates when the penalty term satisfies criterion~\eqref{eq:37}. Therefore, as $\eta$ increases,~\eqref{eq:penalty} will be ultimately satisfied with the desired accuracy $\varepsilon_1$. For the inner loop, $\{\tau,\varepsilon,u,v, \{\mathbf{A}_i\}_{i=1,2}\}$ are jointly optimized by iteratively solving the relaxed problem~\eqref{eq:optimization_problem3} for the given penalty factor. The objective function value of the relaxed version of~\eqref{eq:optimization_problem3} is non-decreasing with each iteration, and its optimal value is bounded. Therefore, the developed penalty-based iterative algorithm is guaranteed to converge to
a stationary point of the original problem~\eqref{eq:optimization_problem}~\cite{dinh2010local}.

The computational complexity of the proposed \textbf{Algorithm 1} is analyzed as follows. 
The complexity for the initialization of the channel correlation matrix $\mathbf{H}_{\text{cor}}$ is given by $O\left((2M)^2\right)$, when $M$-point Gauss-Legendre quadrature is adopted. Note that, referring to~\cite{10910020}, the result obtained using the Gauss-Legendre quadrature method is sufficiently accurate for calculating the channel correlations when $M=20$. 
Moreover, if the interior point method is employed, the computational complexity for solving the relaxed problem (45) is $\mathcal{O}\left(2^{6}\right)$~\cite{luo2010semidefinite}. Therefore, the overall computational complexity of \textbf{Algorithm 1} is $\mathcal{O}\left((2M)^2+2^{6}I_{\mathrm{out}}I_{\mathrm{inn}}\right)$, where $I_{\mathrm{out}}$ and $I_{\mathrm{inn}}$ denote the number of outer and inner iterations required for convergence, respectively. 

Note that problem~\eqref{eq:optimization_problem} can also be solved using the Fourier-based discretization approach proposed in~\cite{10158997}. For the ease of comparison, we also provide the computational complexity analysis of the Fourier-based approach here. Specifically, the complexity for the Fourier transform of the channel is given by $\mathcal{O}\left(2N_FM^2\right)$, where $N_F:=(2N_x+1)(2N_y+1)(2N_z+1)$ is the total number of the reserved Fourier expansion items with $N_x$, $N_y$ and $N_z$ being the numbers of the reserved expansion items on the $x$-, $y$-, and $z$-axis, respectively~\cite{10158997}. Moreover, the computational complexity for the optimization of the Fourier coefficients is given by $\mathcal{O}\left(I_{\mathrm{out}}I_{\mathrm{inn}}N_F^{6}\right)$, when \textbf{Algorithm 1} is adopted. Note that $N_F$ increases significantly with the aperture size and frequency. For instance, considering the case where the CAPA is with the size $L_x=L_y=0.5~\mathrm{m}$, the number of Fourier expansion items $N_F$ is $81$, $729$ and $2601$ when frequency is set as $2.4~\mathrm{GHz}$, $7.8~\mathrm{GHz}$ and $15~\mathrm{GHz}$, respectively. Since $N_F \gg 2$, the proposed channel subspace-based approach effectively reduces the computational complexity compared to the Fourier-based approach. The performance comparison between two approaches will be discussed in Section V.

\section{Low-Complexity Two-Stage ZF-MRT Approach}
Note that the channel subspace-based approach requires double-loop iterations for updating the weighting factors $\left\{\mathbf{a}_i\right\}_{i=1,2}$. For further reducing the computational complexity, in this section, we propose a two-stage source current patters design approach. Specifically, with the aim of completely eliminating the interference of the AN and enhancing the information signal strength at the IR, we design $J_\mathrm{A}\left(\mathbf{s}\right)$ and $J_\mathrm{I}\left(\mathbf{s}\right)$ as the ZF and the MRT beamformers, respectively. Subsequently, the one-dimensional search method is invoked for solving the remaining power allocation problem. 

\subsection{ZF-MRT-Based Source Current Patterns Design}
For the beamforming design in conventional MIMO systems, the ZF beamformer is obtained by computing the pseudoinverse of the users' channel vectors. However, this approach is not applicable to CAPA systems, where the channel vectors are not with finite dimensions but rather continuous functions. To address this challenge, we derive a closed-form ZF solution for the source current patterns design based on the channel correlation matrix $\mathbf{H}_{\text{cor}}$ as follows. 

\textbf{Proposition 1.} Define the inverse matrix of $\mathbf{H}_{\text{cor}}$ as 
\begin{align}
\mathbf{H}_{\text{cor}}^{-1} =
\begin{bmatrix}
u_\mathrm{I,I} & u_\mathrm{I,E} \\
u_\mathrm{E,I} & u_\mathrm{E,E}
\end{bmatrix}.
\end{align}
Then, the ZF beamformer $J_{A}(\mathbf{s})$ is given by
\begin{align}
\label{eq:JA_ZF}
J_\mathrm{A}\left(\mathbf{s}\right)=\sqrt{\rho_\mathrm{A}}J_\mathrm{A}^{\mathrm{ZF}}\left(\mathbf{s}\right),
\end{align}
\begin{align}
\label{eq:JA_ZF2}
J_\mathrm{A}^{\mathrm{ZF}}\left(\mathbf{s}\right)=u_\mathrm{I,E}H_\mathrm{I}\left(\mathbf{s}\right) + u_\mathrm{E,E}H_\mathrm{E}\left(\mathbf{s}\right),
\end{align}
where $\rho_\mathrm{A}$ is the power scaling factor.
\begin{proof}
According to~\eqref{eq:JA_ZF}, the electric field generated by the AN at the IR is given by
\begin{align}
& \int_{\mathcal{S}_\mathrm{T}}H_\mathrm{I}^*\left(\mathbf{s}\right)J_\mathrm{A}\left(\mathbf{s}\right)d\mathbf{s}\nonumber\\
& = \sqrt{\rho_\mathrm{A}}\left(\int_{\mathcal{S}_{\mathrm{T}}}u_\mathrm{I,E}H_{I}\left(\mathbf{s}\right)H_{I}^{*}\left(\mathbf{s}\right)+u_\mathrm{E,E}H_{E}\left(\mathbf{s}\right)H_{I}^{*}\left(\mathbf{s}\right)d\mathbf{s}\right)\nonumber\\
& = \sqrt{\rho_\mathrm{A}}\bar{\mathbf{h}}_\mathrm{I}^T\mathbf{u}_\mathrm{E},
\end{align}
where $\bar{\mathbf{h}}_\mathrm{I}=[h_\mathrm{I,I},h_\mathrm{I,E}]^T$ is the first column of $\mathbf{H}_{\text{cor}}^T$ and $\mathbf{u}_\mathrm{E}=[u_\mathrm{I,E},u_\mathrm{E,E}]^T$ is the second column of $\mathbf{H}_{\text{cor}}^{-1}$. Since $\mathbf{H}_{\text{cor}}\mathbf{H}_{\text{cor}}^{-1}=\mathbf{I}_{2}$, we must have
\begin{subequations}
\label{eq:49}
\begin{equation}
\label{eq:49a}
\bar{\mathbf{h}}_\mathrm{I}^T\mathbf{u}_\mathrm{E}=0,
\end{equation}
\begin{equation}
\label{eq:49b}
\bar{\mathbf{h}}_\mathrm{E}^T\mathbf{u}_\mathrm{E}=1.
\end{equation}
\end{subequations}
Therefore, the interference of the AN is completely eliminated at the IR. 
\end{proof}

Based on the closed-form $J_{\text{A}}\left(\mathbf{s}\right)$ derived in \textbf{Proposition 1}, we can obtain the electric field generated by the AN at the Eve as
\begin{align}
&\int_{\mathcal{S}_\mathrm{T}}H_\mathrm{E}^{*}\left(\mathbf{s}\right)J_\mathrm{A}\left(\mathbf{s}\right)d\mathbf{s} = \sqrt{\rho_\mathrm{A}}\bar{\mathbf{h}}_\mathrm{E}^T\mathbf{u}_\mathrm{E} =\sqrt{\rho_\mathrm{A}}.
\end{align}

Next, we design the beamformer $J_\mathrm{I}\left(\mathbf{s}\right)$ using the MRT beamforming scheme. For conventional MIMO systems with spatially discrete antenna arrays, the MRT beamformer is the conjugate transpose of the user's channel vector. Similarly, the source current pattern $J_\mathrm{I}\left(\mathbf{s}\right)$ is given by
\begin{align}
\label{eq:JI_MRT}
J_\mathrm{I}(\mathbf{s})=\sqrt{\rho_\mathrm{I}}J_\mathrm{I}^{\mathrm{MRT}}(\mathbf{s}),
\end{align}
\begin{align}
\label{eq:JI_MRT2}
J_\mathrm{I}^{\mathrm{MRT}}(\mathbf{s})=&{H_\mathrm{I}(\mathbf{s})},
\end{align}
where $\rho_\mathrm{I}$ is the power scaling factor. 
%It is interesting to find that the ZF and the MRT source current patterns also lie within the channel subspace spanned by $H_\mathrm{I}\mathbf{s}$ and $H_\mathrm{E}\mathbf{s}$. 

Subsequently, we need to address the remaining power allocation problem. Denote the power allocated to the information and AN signals as $P_{\text{I}}$ and $P_{\text{A}}$, respectively. Then we have 
\begin{align}
\label{eq:rhoA}
\rho_\mathrm{A}=\frac{P_\mathrm{A}}{\int_{\mathcal{S}_\mathrm{T}}\left|J_\mathrm{A}^\mathrm{ZF}\left(\mathbf{s}\right)\right|^2d\mathbf{s}}\overset{(c)}{\operatorname*{\operatorname*{=}}}\frac{P_\mathrm{A}}{\mathbf{u}_\mathrm{E}^H\mathbf{H}_{\text{cor}}\mathbf{u}_\mathrm{E}}=\frac{P_\mathrm{A}}{u_\mathrm{E,E}},
\end{align}
\begin{align}
\label{eq:rhoI}
\rho_\mathrm{I}=\frac{P_\mathrm{I}}{\int_{\mathcal{S}_\mathrm{T}}\left|J_\mathrm{I}^\mathrm{MRT}\left(\mathbf{s}\right)\right|^2d\mathbf{s}}=\frac{P_\mathrm{I}}{h_\mathrm{I,I}}.
\end{align}
In~\eqref{eq:rhoA}, equality $(c)$ is obtained based on the detailed derivations given in~\eqref{eq:57} shown at the top of the next page. Moreover, equality $(d)$ can be proved based on~\eqref{eq:49}, which is detailed as
\begin{figure*}[!h]
    \begin{align}
& \int_{\mathcal{S}_\mathrm{T}}\left|J_\mathrm{A}^\mathrm{ZF}\left(\mathbf{s}\right)\right|^2d\mathbf{s}\nonumber\\
& =\int_{\mathcal{S}_\mathrm{T}}\left(\left|u_\mathrm{I,E}\right|^2H_\mathrm{I}\left(\mathbf{s}\right)H_\mathrm{I}^*\left(\mathbf{s}\right) + u_\mathrm{I,E}u_\mathrm{E,E}^*H_\mathrm{I}\left(\mathbf{s}\right)H_\mathrm{E}^*\left(\mathbf{s}\right)\nonumber+ u_\mathrm{E,E}u_\mathrm{I,E}^*H_\mathrm{E}\left(\mathbf{s}\right)H_\mathrm{I}^*\left(\mathbf{s}\right)d\mathbf{s} + \left|u_\mathrm{E,E}\right|^2H_\mathrm{E}\left(\mathbf{s}\right)H_\mathrm{E}^*\left(\mathbf{s}\right)\right)d\mathbf{s}\nonumber\\
& =\left|u_\mathrm{I,E}\right|^2h_\mathrm{I,I} + u_\mathrm{I,E}u_\mathrm{E,E}^*h_\mathrm{E,I} + u_\mathrm{E,E}u_\mathrm{I,E}^*h_\mathrm{I,E} + \left|u_\mathrm{E,E}\right|^2h_\mathrm{E,E}\nonumber\\
&= \mathbf{u}_\mathrm{E}^H\mathbf{H}_{\text{cor}}\mathbf{u}_\mathrm{E},
    \label{eq:57}
\end{align}
\hrulefill
\end{figure*}
\begin{align}
\mathbf{u}_\mathrm{E}^H\mathbf{H}_{\text{cor}}\mathbf{u}_\mathrm{E} = \left(\mathbf{H}_{\text{cor}}\mathbf{u}_\mathrm{E}\right)^H\mathbf{u}_\mathrm{E} = \mathbf{e}_2^H\mathbf{u}_\mathrm{E} = u_\mathrm{E,E},
\end{align}
where $\mathbf{e}_2=[0,1]^T$.
Accordingly, with the ZF and MRT solutions given in~\eqref{eq:JA_ZF} and~\eqref{eq:JI_MRT}, the SINR at the IR and the Eve can be rewritten with respect to $P_{\text{A}}$ and $P_{\text{I}}$ as
\begin{align}
\bar{\gamma}_\mathrm{I} & =\frac{\rho_\mathrm{I}\left|\int_{\mathcal{S}_{\mathrm{T}}}H_\mathrm{I}^*\left(\mathbf{s}\right)J_\mathrm{I}^{\mathrm{MRT}}\left(\mathbf{s}\right)d\mathbf{s}\right|^{2}}{\rho_\mathrm{A}\left|\int_{\mathcal{S}_{\mathrm{T}}}H_\mathrm{I}^*\left(\mathbf{s}\right)J_\mathrm{A}^{\mathrm{ZF}}\left(\mathbf{s}\right)d\mathbf{s}\right|^{2}+\sigma_\mathrm{I}^{2}}\nonumber\\
& =\frac{\rho_\mathrm{I}h_\mathrm{I,I}}{\rho_\mathrm{A}\left|\bar{\mathbf{h}}_\mathrm{I}^T\mathbf{u}_\mathrm{E}\right|^{2}+\sigma_\mathrm{I}^{2}}\nonumber\\
& = \frac{P_\mathrm{I}h_\mathrm{I,I}}{\sigma_\mathrm{I}^{2}},
\end{align}
and
\begin{align}
\bar{\gamma}_\mathrm{E} & =\frac{\rho_\mathrm{I}\left|\int_{\mathcal{S}_{\mathrm{T}}}H_\mathrm{E}^*\left(\mathbf{s}\right)J_\mathrm{I}^{\mathrm{MRT}}\left(\mathbf{s}\right)d\mathbf{s}\right|^{2}}{\rho_\mathrm{A}\left|\int_{\mathcal{S}_{\mathrm{T}}}H_\mathrm{E}^*\left(\mathbf{s}\right)J_\mathrm{A}^{\mathrm{ZF}}\left(\mathbf{s}\right)d\mathbf{s}\right|^{2}+\sigma_\mathrm{E}^{2}}\nonumber\\
& 
=\frac{\rho_\mathrm{I}\frac{\left|h_\mathrm{E,I}\right|^{2}}{h_\mathrm{I,I}}}{\rho_{A}\left|\bar{\mathbf{h}}_\mathrm{E}^T\mathbf{u}_\mathrm{E}\right|^{2}+\sigma_\mathrm{E}^{2}}\nonumber\\
&
=\frac{P_\mathrm{I}\left|h_\mathrm{E,I}\right|^{2}u_\mathrm{E,E}}{h_\mathrm{I,I}\left(P_\mathrm{A}+u_\mathrm{E,E}\sigma_\mathrm{E}^{2}\right)},
\end{align}
respectively.
As such, $R_{\mathrm{S}}$ in the original problem~\eqref{eq:optimization_problem} can be rewritten as
\begin{align}
\label{eq:RS_ZF}
\bar{R}_{\mathrm{S}} & = \log\left(1+\bar{\gamma}_\mathrm{I}\right) - \log\left(1+\bar{\gamma}_\mathrm{E}\right) \nonumber\\
& =\log\left(1+\frac{P_\mathrm{I}h_\mathrm{I,I}}{\sigma_\mathrm{I}^{2}}\right)\nonumber\\
&-\log\left(1+\frac{P_\mathrm{I}\left|h_\mathrm{E,I}\right|^{2}u_\mathrm{E,E}}{h_\mathrm{I,I}\left(P_\mathrm{A}+u_\mathrm{E,E}\sigma_\mathrm{E}^{2}\right)}\right).
\end{align}

\subsection{Power Allocation}
Then, the original problem~\eqref{eq:optimization_problem} can be reduced to the following power allocation problem:
\begin{subequations}
\label{eq:power allocation}
\begin{equation}
\label{eq:66aSRM_objective}
\max_{P_\mathrm{I},P_\mathrm{A}}
\bar{R}_\mathrm{S}
\end{equation}
\begin{equation}
\label{eq:66bpower_constraint}
    {\rm{s.t.}} \ \ P_\mathrm{I}+P_\mathrm{A}= P_\mathrm{T}.
\end{equation}
\end{subequations}
Here, the previous inequality constraint~\eqref{eq:15power_constraint} is equivalently transformed into the equality constraint~\eqref{eq:66bpower_constraint} in problem~\eqref{eq:power allocation}. To demonstrate it, suppose that there exists an optimal power allocation pair $\left(P_\mathrm{I}, P_\mathrm{A}\right)$ that satisfies $P_\mathrm{I}+P_\mathrm{A}<P_\mathrm{T}$. We can always increase the $P_\mathrm{A}$ until $P_\mathrm{I}+P_\mathrm{A}=P_\mathrm{T}$ without decreasing the secrecy rate. Therefore, the optimal power allocation to problem~\eqref{eq:power allocation} must satisfy $P_\mathrm{I}+P_\mathrm{A}=P_\mathrm{T}$.

Let $P_\mathrm{A}=P_\mathrm{T}-P_\mathrm{I}$, problem~\eqref{eq:66bpower_constraint} can be further simplified to
\begin{align}
\label{eq:RS_PI}
\max_{0\leq P_\mathrm{I} \leq P_\mathrm{T}}
\log\left(1+qP_\mathrm{I}\right) - \log\left(1+\frac{tP_\mathrm{I}}{P_\mathrm{T}-P_\mathrm{I}+c}\right),
\end{align}
where $q=\frac{h_\mathrm{I,I}}{\sigma_\mathrm{I}^{2}}$,$t=\frac{\left|h_\mathrm{E,I}\right|^{2}u_\mathrm{E,E}}{h_\mathrm{I,I}}$ and $c=u_\mathrm{E,E}\sigma_\mathrm{E}^{2}$. Problem~\eqref{eq:RS_PI} be directly solved using the classical one-dimensional search method.

\begin{algorithm}[h]
\renewcommand{\algorithmicrequire}{\textbf{Input:}}
\renewcommand{\algorithmicensure}{\textbf{Output:}}
\caption{Proposed Two-Stage ZF-MRT Algorithm}
\label{alg:ZF-MRT}
    \begin{algorithmic}[1]
    \Statex \underline{\textbf{Stage 1}: ZF and MRT beamformers design}
    \State Calculate the channel correlation matrix $\mathbf{H}_{\text{cor}}$.
    \State Calculate the inverse of the matrix $\mathbf{H}_{\text{cor}}$ and obtain $u_\mathrm{E,E}$.
    \State Obtain the ZF beamformer with \eqref{eq:JA_ZF} and \eqref{eq:JA_ZF2}.
    \State Obtain the MRT beamformer with \eqref{eq:JI_MRT} and \eqref{eq:JI_MRT2}.
    \Statex \underline{\textbf{Stage 2}: Power allocation}
    \State Calculate the coefficients $q, t$, and $c$, i.e., $q=\frac{h_\mathrm{I,I}}{\sigma_\mathrm{I}^{2}}$,$t=\frac{\left|h_\mathrm{E,I}\right|^{2}u_\mathrm{E,E}}{h_\mathrm{I,I}}$, $c=u_\mathrm{E,E}\sigma_\mathrm{E}^{2}$.
    \State Solve problem~\eqref{eq:RS_PI} using the one-dimensional search method.
    \end{algorithmic}
\end{algorithm}

\subsection{Complexity Analysis}
The proposed two-stage ZF-MRT algorithm is summarized in \textbf{Algorithm 2}. The computational complexity of the proposed ZF-MRT approach is analyzed as follows. The complexity for the calculation of the channel correlation matrix $\mathbf{H}_{\text{cor}}$ is given by $O\left((2M)^2\right)$. Moreover, the computational complexity also arises from calculating the inverse of the matrix $\mathbf{H}_{\text{cor}}$, which has a worst-case complexity of $O\left(2^3\right)$. Additionally, the computational complexity of the one-dimensional search for obtaining the optimal $P_\mathrm{I}$ is $\mathcal{O}\left(P_\mathrm{T}/\varepsilon\right)$, where $\varepsilon$ denotes the accuracy tolerance value. Therefore, the total computational complexity of the proposed two-stage ZF-MRT algorithm is $O\left((2M)^2+2^3+P_\mathrm{T}/\varepsilon\right)$. Recall that the complexity of the channel subspace-based approach proposed in Section III is given by $\mathcal{O}\left((2M)^2+2^{6}I_{\mathrm{out}}I_{\mathrm{inn}}\right)$. It can be observed that the computational complexity of the ZF-MRT approach is lower than that of the channel subspace-based approach, given that the double-loop iterations are efficiently avoided.  

\section{Numerical Results}
In this section, numerical results are provided to evaluate the effectiveness of the channel subspace-based and the ZF-MRT approaches for the secure beamforming design in the CAPA-based communications system. 

\subsection{Simulation Setup and Baselines}
It is assumed that the transmit CAPA lies within the $xy$-plane and is centered at the origin of the coordinate system, defined as
\begin{align}
\mathcal{S}_{\mathrm{T}}=\left\{[s_x,s_y,0]^T 
     \Big||s_x|\leq\frac{L_x}{2},|s_y|\leq\frac{L_y}{2}\right\}.
\end{align}
The transmit CAPA has a square shape with the area of $A_\mathrm{T}=0.25~\mathrm{m}^{2}$, i.e., $L_x=L_y=\sqrt{A_\mathrm{T}}$. The system contains one IR and one Eve, which are located within the following region
\begin{align}
\mathcal{R}=\left\{[r_x,r_y,r_z]^T\Bigg|
\begin{array}
{c}|r_x|\leq U_x,|r_y|\leq U_y, \\
U_{z,\min}\leq r_z\leq U_{z,\max}
\end{array}\right\},
\end{align}
where $U_{x}=U_{y}=5~\mathrm{m}$, $U_{z,\min}=15~\mathrm{~m}$ and $U_{z,\max}=30~\mathrm{~m}$. Without loss of generality, assume that the polarization directions of the IR and the Eve are aligned along the y-axis, i.e., $\hat{\mathbf{u}}_\mathrm{I}=\hat{\mathbf{u}}_\mathrm{E}=\hat{\mathbf{u}}_{y}=[0,1,0]^{T}$. Unless stated otherwise, the signal frequency of the source current and the intrinsic impedance are set as $f=2.4~\mathrm{GHz}$ {and $\eta=120\pi~\Omega$}, respectively. The maximum transmit power is set to $P_{\mathrm{T}}=100~\mathrm{mA}^{2}$, while the noise power for the IR and the Eve is configured as $\sigma_\mathrm{I}^{2}=\sigma_\mathrm{E}^{2}=5.6\times10^{-3}~\mathrm{V}^{2}/\mathrm{m}^{2}$. The number of sample points of the Gauss-Legendre quadrature for calculating all integrals in the simulation is set to $M=20$. All simulation results are obtained by averaging over 200 independent random channel realizations unless otherwise specified.

To fully demonstrate the significant advantages of the proposed schemes in improving the CAPA system’s secrecy performance, we consider the following benchmark schemes. 

\begin{itemize}
\item \textbf{Discrete MIMO}: In this case, the planar spatially discrete array is exploited. The continuous surface $\mathcal{S}_{\mathrm{T}}$ is occupied with discrete antennas, spaced with $d=\frac{\lambda}{2}$. The effective aperture area of each antenna is given by $A_d=\frac{\lambda^2}{4\pi}$. The coordinate position of the $(n_x,n_y)$-th discrete antenna is given by 
\begin{align}
\mathbf{\bar{s}}_{n_x,n_y}=\left[\left(n_x-1\right)d-\frac{L_x}{2},\left(n_y-1\right)d-\frac{L_y}{2},0\right]^T.    
\end{align}
Therefore, the total number of discrete antennas is $N_d=\lceil\frac{L_x}{d}\rceil\times\lceil\frac{L_y}{d}\rceil$. The channel between the $(n_x,n_y)$-th discrete antenna and the IR/Eve can be calculated as
\begin{subequations}
\begin{equation}
h_{\left(n_x,n_y\right),\mathrm{I}}=\sqrt{A_d}\hat{\mathbf{u}}_\mathrm{I}^T\mathbf{G}\left(\mathbf{r}_\mathrm{I},\mathbf{\bar{s}}_{n_x,n_y}\right)\hat{\mathbf{u}}_y,
\end{equation}
\begin{equation}
h_{\left(n_x,n_y\right),\mathrm{E}}=\sqrt{A_d}\hat{\mathbf{u}}_\mathrm{E}^T\mathbf{G}\left(\mathbf{r}_\mathrm{E},\mathbf{\bar{s}}_{n_x,n_y}\right)\hat{\mathbf{u}}_y.
\end{equation}
\end{subequations}

Note that the optimal beamforming vectors for the information and AN signals also lie in the channel subspace spanned by the discrete channels, for which the proof can be given similar to that in \textbf{Appendix A}. As such, we apply the channel subspace-based approach for the beamforming design in the discrete MIMO case as well, which is with much lower complexity compared to the conventional beamforming methods, such as the iterative algorithm~\cite{6587993}. Moreover, the method combining the ZF and the MRT schemes is also applied to the discrete MIMO case, which is regarded as a separate benchmark.

\item \textbf{Fourier-based approach}: In this case, the state-of-the-art Fourier-based approach is adopted~\cite{10158997}. The main idea of the Fourier-based approach is to approximate source current patterns using a finite number of Fourier series. Specifically, the source current patterns can be expressed as
\begin{subequations}
\begin{equation}
J_\mathrm{I}\left(\mathbf{s}\right)\approx\sum_{\mathbf{n}=-\mathbf{N}}^\mathbf{N}v_{\mathrm{I},\mathbf{n}}\Phi_\mathbf{n}\left(\mathbf{s}\right),
\end{equation}
\begin{equation}
J_\mathrm{A}\left(\mathbf{s}\right)\approx\sum_{\mathbf{n}=-\mathbf{N}}^\mathbf{N}v_{\mathrm{A},\mathbf{n}}\Phi_\mathbf{n}\left(\mathbf{s}\right),
\end{equation}
\end{subequations}
where $\mathbf{n}=[n_x,n_y,n_z]^T$, and the sum notation is defined as $\sum_{\mathbf{n}=-\mathbf{N}}^{\mathbf{N}}\triangleq \sum_{n_x=-N_x}^{N_x}\sum_{n_y=-N_y}^{N_y}\sum_{n_z=-N_z}^{N_z}$. $v_{\mathrm{I},\mathbf{n}}$ ($v_{\mathrm{A},\mathbf{n}}$) and $\Phi_\mathbf{n}(\mathbf{s})$ denote Fourier coefficients and orthonormal Fourier basis functions, respectively. The resultant Fourier coefficients optimization problem is solved by employing $\textbf{Algorithm 1}$ described in Section III-B.

\item \textbf{MRT approach}: In this case, the MRT approach described in Section IV is adopted to design both the source current patterns $J_\mathrm{I}\left(\mathbf{s}\right)$ and $J_\mathrm{A}\left(\mathbf{s}\right)$. Specifically, $J_\mathrm{I}\left(\mathbf{s}\right)$ and $J_\mathrm{A}\left(\mathbf{s}\right)$ are given by
\begin{subequations}
\begin{equation}
J_\mathrm{I}\left(\mathbf{s}\right)=\sqrt{\frac{P_\mathrm{I}}{\int_{\mathcal{S}_\mathrm{T}}\left|H_\mathrm{I}(\mathbf{s})\right|^2d\mathbf{s}}}H_\mathrm{I}\left(\mathbf{s}\right),
\end{equation}
\begin{equation}
J_\mathrm{A}\left(\mathbf{s}\right)=\sqrt{\frac{P_\mathrm{A}}{\int_{\mathcal{S}_\mathrm{T}}\left|H_\mathrm{E}\left(\mathbf{s}\right)\right|^2d\mathbf{s}}}H_\mathrm{E}\left(\mathbf{s}\right).
\end{equation}
\end{subequations}
Then, the remaining power allocation problem is solved using the one-dimensional search.
\end{itemize}

\begin{table}[!t]
\renewcommand\arraystretch{1.5}
\caption{Comparison of average CPU runtime.}%title
\centering
\begin{tabular}{c|cc|cc}% four columns
\toprule[1pt] %change the first line to \toprule
\multirow{2}*{Frequency} & \multicolumn{2}{c|}{$A_\mathrm{T}=0.25~\mathrm{m}^{2}$}  & \multicolumn{2}{c}{$A_\mathrm{T}=0.5~\mathrm{m}^{2}$} \\
& Proposed & Fourier & Proposed & Fourier\\
\cline{1-5} %change the second line to midrule
2.4~GHz & 1.671~s & 21.92~s & 1.879~s & 50.21~s\\
7.8~GHz & 1.934~s & 174.57~s & 1.937~s & 532.98~s\\
\bottomrule[1pt] %change the third line to bottomrule
\end{tabular}
\label{tab:cpu_runtime_comparison}
\end{table}

For improving the convergence speed of \textbf{Algorithm 1}, we utilize the low-complexity ZF-MRT scheme proposed in Section IV for the initialization of the weighting vectors $\mathbf{a}_{1}$ and $\mathbf{a}_{2}$. Specifically, according to~\eqref{eq:JA_ZF2} and~\eqref{eq:JI_MRT2}, $\mathbf{a}_{1}$ and $\mathbf{a}_{2}$ are given by 
\begin{align}
\label{eq:initialization}
\mathbf{a}_{1}^{(0)}=\mathbf{e}_1,
\mathbf{a}_{2}^{(0)}=\frac{\mathbf{u}_\mathrm{E}}{\|\mathbf{u}_\mathrm{E}\|}.
\end{align}
Then, by  substituting~\eqref{eq:initialization} into~\eqref{eq:auxiliary_variables_ep} and~\eqref{eq:auxiliary_variables_u}, the initial auxiliary variables $\varepsilon^{(0)}, u^{(0)}$ can be obtained as
\begin{align}
\varepsilon^{\left(0\right)}=\operatorname{ln}\left(\mathrm{Tr}\left(\mathbf{H}_\mathrm{I}\mathbf{A}_2^{\left(0\right)}\right) + \sigma_\mathrm{I}^2\right),
\end{align}
\begin{align}
u^{\left(0\right)}=\operatorname{ln}\left(\mathrm{Tr}\left(\mathbf{H}_\mathrm{E}\mathbf{A}_1^{\left(0\right)}\right) + \mathrm{Tr}\left(\mathbf{H}_\mathrm{E}\mathbf{A}_2^{\left(0\right)}\right)+ \sigma_\mathrm{E}^2\right),
\end{align}
where $\mathbf{A}_1^{\left(0\right)}=\mathbf{a}_1^{\left(0\right)}\left(\mathbf{a}_1^{\left(0\right)}\right)^H$ and $\mathbf{A}_2^{(0)}=\mathbf{a}_2^{\left(0\right)}\left(\mathbf{a}_2^{(0)}\right)^H$.

\begin{figure}[t]
	\centering
	\includegraphics[scale=0.6]{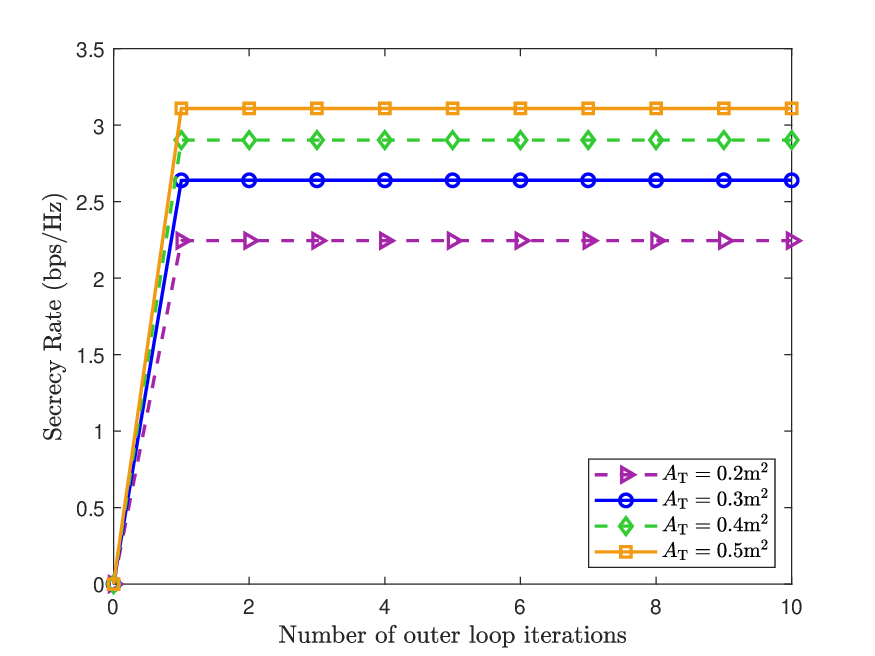}
	\caption{Convergence of the proposed algorithm.}
	\label{fig:iteration_5} 
\end{figure}

\begin{figure}[t]
	\centering
	\includegraphics[scale=0.6]{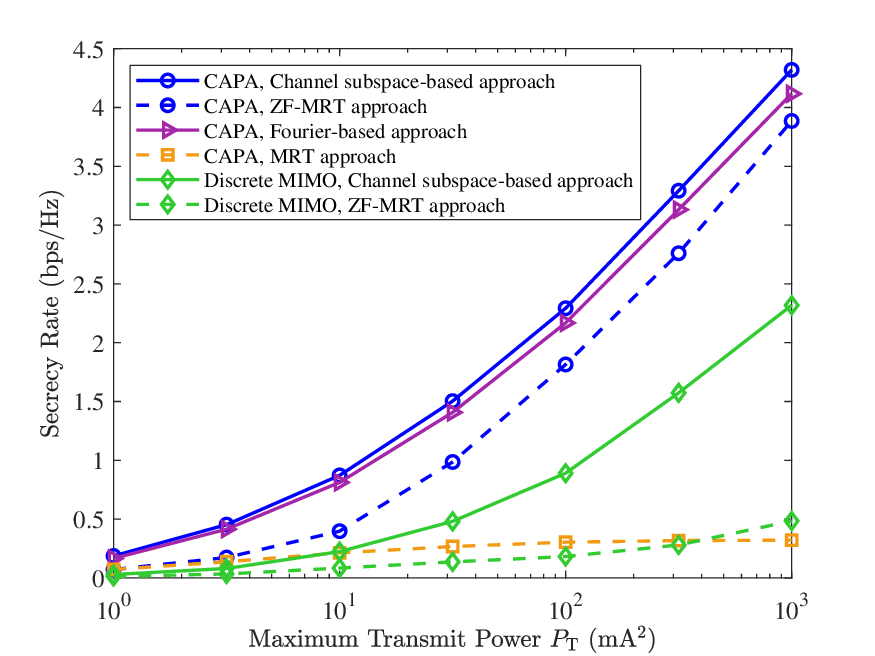}
	\caption{Secrecy rate versus maximum transmit power.}
	\label{fig:PT_new2} 
\end{figure}

\subsection{Convergence and Complexity of Algorithms 1}
 Fig.~\ref{fig:iteration_5} illustrates the convergence behavior of the proposed \textbf{Algorithm 1} under different aperture sizes. It can be seen that the algorithm only requires one iteration for converging to a stable value. Table~\ref{tab:cpu_runtime_comparison} compares the CPU runtime consumed by the proposed channel subspace-based and the Fourier-based approaches using MATLAB R2023b on an Intel i9-13980HX processor. From this table, we can observe that, for the proposed channel subspace-based approach, the CPU runtime increases marginally with the aperture size and frequency. For example, as the aperture size and frequency increase from $0.25~\mathrm{m}^{2}$ and $2.4~\mathrm{GHz}$ to $0.5~\mathrm{m}^{2}$ and $7.8~\mathrm{GHz}$, the CPU runtime consumed by the proposed channel subspace-based approach remains consistently low. This is because, the dimension of the optimization variables are independent of both the aperture size and the frequency, resulting in nearly unchanged complexity. In contrast, the CPU runtime consumed by the Fourier-based approach increases dramatically with the aperture size and frequency. This is due to the fact that, as the aperture size and frequency increase, the required number of Fourier basis functions $N_{\mathrm{F}}$ will increase significantly, which results in much higher computational complexity. These observations are consistent with the computational complexity analysis in Section III-D.

\begin{figure}[t]
	\centering
	\includegraphics[scale=0.6]{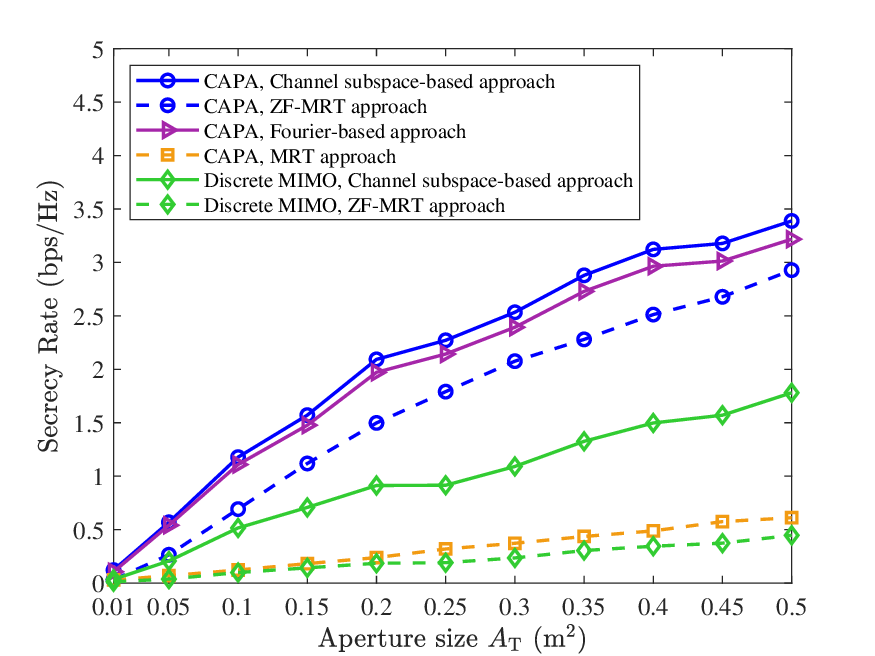}
	\caption{Secrecy rate versus aperture size.}
	\label{fig:AT_new2} 
\end{figure}

\begin{figure}[t]
	\centering
	\includegraphics[scale=0.6]{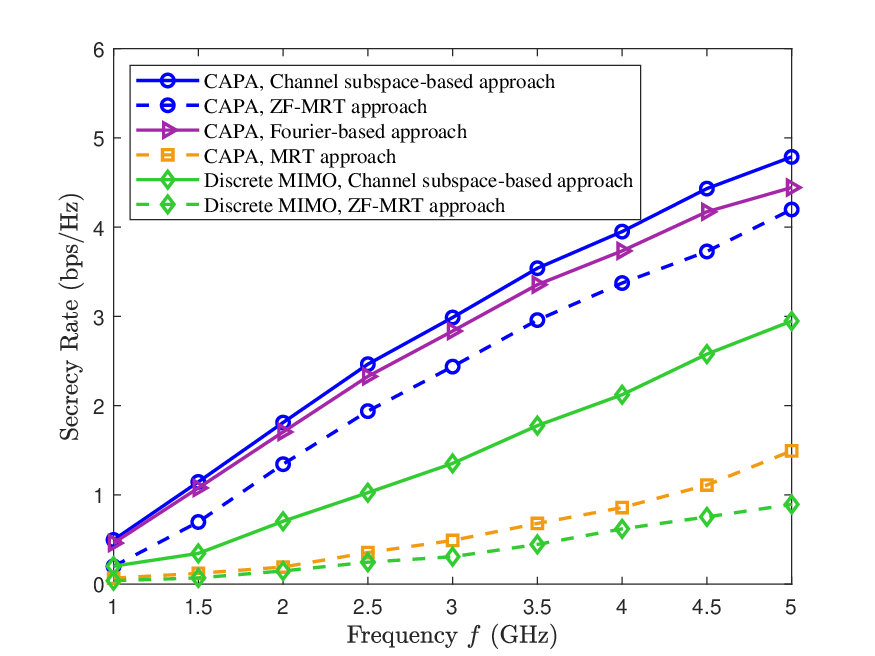}
	\caption{Secrecy rate versus frequency.}
	\label{fig:F_new2} 
\end{figure}

\subsection{Secrecy Rate Versus Maximum Transmit Power}
In Fig.~\ref{fig:PT_new2}, we compare the secrecy rate of difference schemes versus the BS maximum transmit power $P_{\text{T}}$. It can be seen that the CAPA significantly outperforms the conventional discrete MIMO, when the proposed channel subspace-based scheme or the ZF-MRT scheme is applied. For example, when $P_{\mathrm{T}}=10^{3}~\mathrm{mA}^{2}$, the proposed channel subspace-based scheme and the ZF-MRT scheme achieve around $87\%$ and $69\%$ improvement of the secrecy rate over the discrete MIMO, respectively. This is expected, as the CAPA brings in higher spatial DoFs with the continuous source current distributed across the aperture. Regarding the effectiveness of the proposed beamforming methods, one can first observe that the proposed channel subspace-based approach achieves higher secrecy rate compared to the Fourier-based approach. This is due to the fact that, the continuous patterns approximation with Fourier basis functions is effectively avoided in the channel subspace-based approach. It is also observed that, for the CAPA, the ZF-MRT attains prominent performance gain compared to the MRT benchmark, and realizes close performance to the Fourier-based approach with the increment of $P_{\text{T}}$, which implies the effectiveness of the ZF-MRT scheme under large SNR scenarios.     

	\begin{figure}[t]
		\centering
        \label{fig:ap1}
		\subfigure[Channel subspace-based approach.]{
        \label{fig:subspace1}
			\includegraphics[scale=0.6]{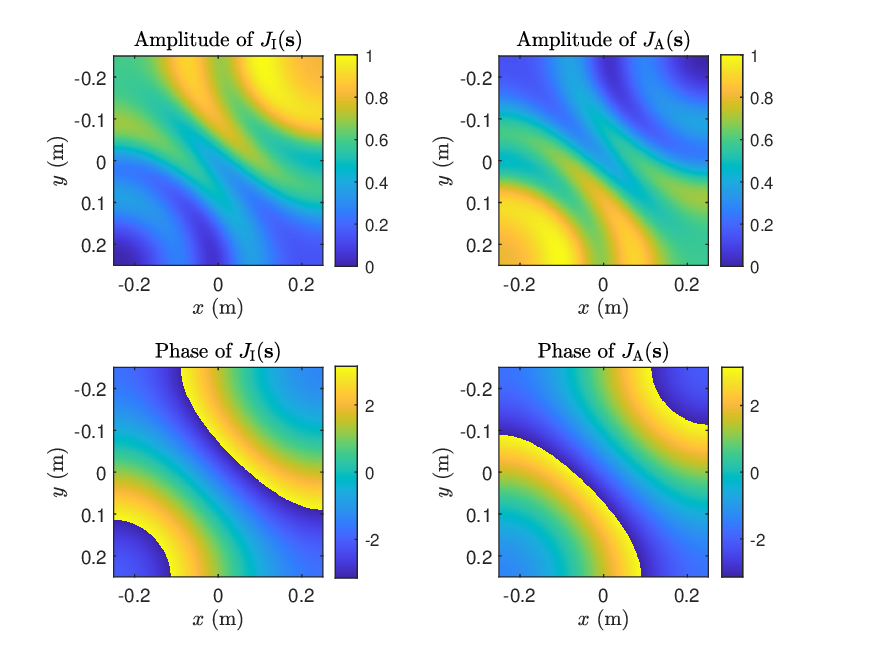}}
		\subfigure[ZF-MRT approach.]{
        \label{fig:ZF-MRT1}
			\includegraphics[scale=0.6]{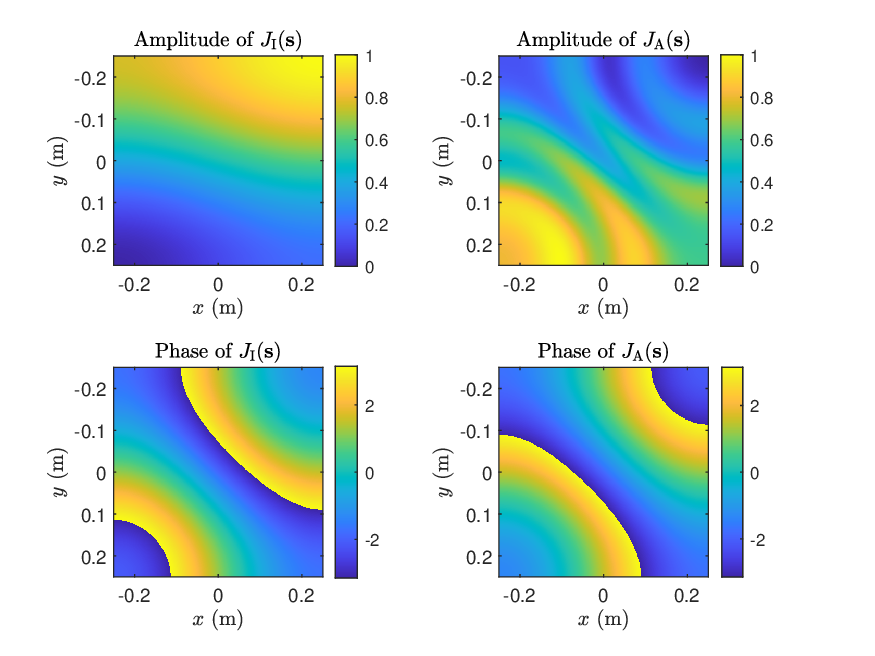}}
		\caption{Amplitudes and phases of source current patterns, where the IR and the Eve are located at $\mathbf{r}_{\text{I}}=[5,-5,20]^T$ and $\mathbf{r}_{\text{E}}=[-5,5,20]^T$, respectively.} 
	\end{figure}

	\begin{figure}[t]
		\centering
        \label{fig:ap3} 
		\subfigure[Channel subspace-based approach.]{
        \label{fig:subspace2}
			\includegraphics[scale=0.6]{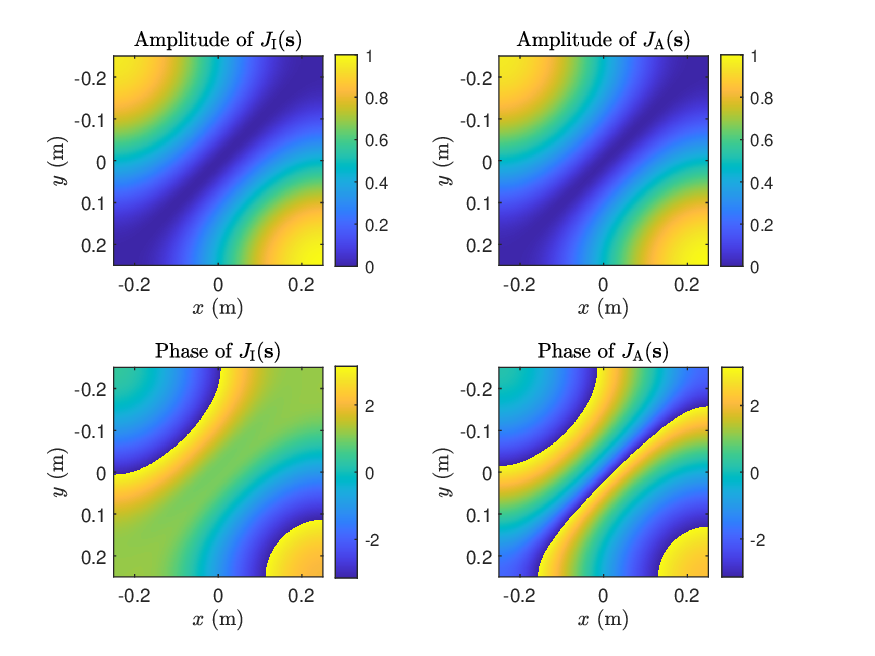}}
		\subfigure[ZF-MRT approach.]{
        \label{fig:ZF-MRT2}
			\includegraphics[scale=0.6]{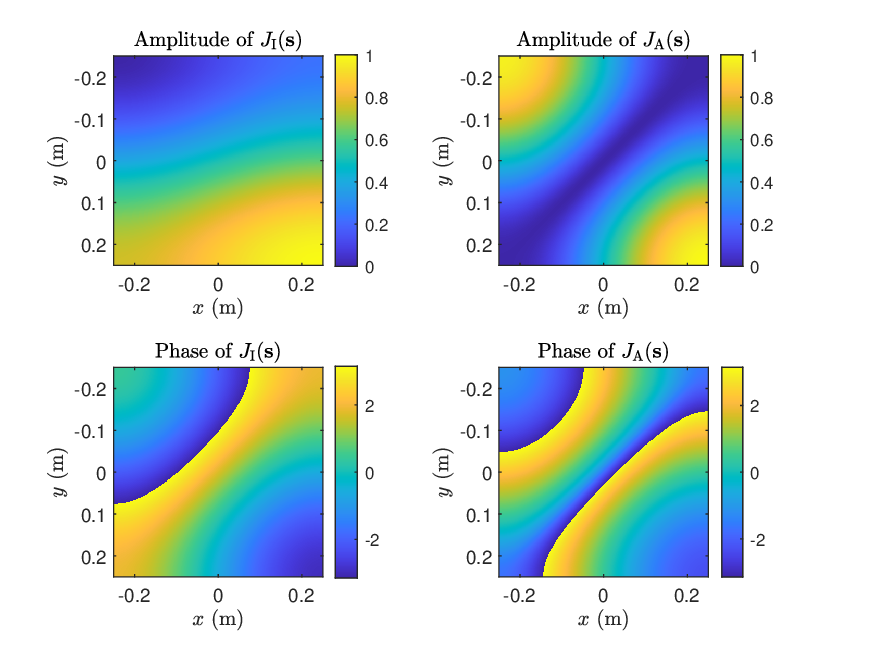}}
		\caption{Amplitudes and phases of source current patterns, where the IR and the Eve are located at $\mathbf{r}_{\text{I}}=[4,4,20]^T$ and $\mathbf{r}_{\text{E}}=[5,5,20]^T$, respectively.}
	\end{figure}

\subsection{Secrecy Rate Versus CAPA Aperture Sizes}
Fig.~\ref{fig:AT_new2} depicts the secrecy rate versus the aperture size $A_{\mathrm{T}}$ of the CAPA. One can observe that the secrecy rate achieved by all schemes increases with larger $A_{\mathrm{T}}$. This can be explained as follows. The increment of the aperture size brings in higher spatial DoFs for the CAPA beamforming, thereby improving the secrecy rate. Similarly, the increment of the aperture size means larger number of antennas for conventional discrete MIMO, which also leads to higher beamforming gain. It is also interesting to find that, as the aperture size increases, the improvement of the secrecy rate achieved by the CAPA is more pronounced compared to the discrete MIMO benchmark. For example, when the channel subspace-based beamforming approach is applied, the CAPA can improve the secrecy rate by around $149\%$ and $90\%$ over the discrete MIMO under $A_{\text{T}}=0.25~\mathrm{m}^{2}$ and $A_{\text{T}}=0.5~\mathrm{m}^{2}$, respectively. This is expected, as the continuous source current can fully exploit the spatial DoFs, and thereby bringing in higher performance gain with larger aperture size. 

\subsection{Secrecy Rate Versus Operating Frequencies}
In Fig.~\ref{fig:F_new2}, we investigate the impact of the operating frequency $f$ on the achievable secrecy rate. We can observe that, with the increment of the frequency, the secrecy rate achieved by all schemes enhances. This is consistent with the conclusion provided in~\cite{bjornson2024enabling} that, the received power increasing quadratically as the wavelength $\lambda$ shrinks.
These results highlights the importance of leveraging higher frequency to enhance security performance in future wireless systems. Notably, the computational complexity of the Fourier-based approach increases with higher frequency, given the larger number of Fourier basis functions $N_{\text{F}}$. Therefore, the superiority of the proposed beamforming schemes is more pronounced under higher operating frequency.  

\subsection{Source Current Patterns}
Fig.~\ref{fig:subspace1} and Fig.~\ref{fig:ZF-MRT1} demonstrate the amplitudes and phases of source current patterns $J_\mathrm{I}(\mathbf{s})$ and $J_\mathrm{A}(\mathbf{s})$ obtained using the proposed channel subspace-based approach and ZF-MRT approach, respectively, where the IR and the Eve are located at $\mathbf{r}_{\text{I}}=[5,-5,20]^T$ and $\mathbf{r}_{\text{E}}=[-5,5,20]^T$, respectively. One can first observe that, as the IR and the Eve are located symmetrically, the amplitudes and phases of source current patterns $J_\mathrm{I}(\mathbf{s})$ and $J_\mathrm{A}(\mathbf{s})$ obtained using the channel subspace-based approach are symmetrically distributed. This indicates that the EM waves carrying the information and AN signals are steered toward the IR and the Eve, respectively. It is also seen that, the amplitudes and phases of source current patterns $J_\mathrm{I}(\mathbf{s})$ and $J_\mathrm{A}(\mathbf{s})$ obtained using the ZF-MRT approach are similar to that obtained by the channel subspace-based approach, which underscores the effectiveness of the ZF-MRT approach when the IR and the Eve are located far apart. 

Fig.~\ref{fig:subspace2} and Fig.~\ref{fig:ZF-MRT2} further depict the amplitudes and phases of source current patterns $J_\mathrm{I}(\mathbf{s})$ and $J_\mathrm{A}(\mathbf{s})$, respectively, when the IR and the Eve are close to each other, located at $\mathbf{r}_{\text{I}}=[4,4,20]^T$ and $\mathbf{r}_{\text{E}}=[5,5,20]^T$, respectively. It is observed that, $J_\mathrm{I}(\mathbf{s})$ obtained by the MRT scheme differs much from that obtained by the channel subspace-based approach. This is expected, as the MRT can lead to severe information leakage when the IR is located close to the Eve. Therefore, the performance of the MRT scheme becomes deteriorated. On the contrary, $J_\mathrm{A}(\mathbf{s})$ obtained by the ZF-MRT approach is similar to that obtained by the channel subspace-based approach. This implies that, the source current pattern of the AN signal is preferred to be designed to be orthogonal to the channel of the IR, so as to suppress the interference to the IR. 

\section{Conclusions}
A secure communication framework in CAPA systems was investigated, where the BS equipped with a CAPA simultaneously transmitted the information signal and the AN for the jamming purpose. To effectively address the non-convex integral-based functional programming problem for optimizing the source current patterns, a channel subspace-based beamforming scheme was proposed. Specifically, by exploiting the subspace spanned by all users’ channel responses, the original problem is equivalently converted to a channel-subspace weighting factors optimization problem, which was effectively solved by invoking the SCA method. To further reduce the computational complexity, a two-stage source current
patterns design scheme was proposed. Specifically, the closed-form beamformers were derived based on the ZF and MRT schemes, which was followed by the power allocation addressed with the one-dimensional search. It was shown that the CAPA brought significant secrecy rate gain compared to the discrete MIMO. Moreover, the superiority of the proposed channel subspace-based approach was demonstrated in both secrecy performance and the computational complexity compared to the state-of-the-art Fourier-based approach.  

\begin{appendices} 
\section{PROOF OF LEMMA 1}
First, we prove~\eqref{eq:linerJI} with its converse-negative proposition. Specifically, suppose there exists an optimal solution $J_\mathrm{I}\left(\mathbf{s}\right)^{\prime}$ of problem~\eqref{eq:optimization_problem} that does not lie within the subspace spanned by $H_\mathrm{I}\left(\mathbf{s}\right)$ and $H_\mathrm{E}\left(\mathbf{s}\right)$. In other words, we have
\begin{align}
J_\mathrm{I}^{\prime}\left(\mathbf{s}\right)=\alpha_{11}^{\prime}H_\mathrm{I}\left(\mathbf{s}\right) + \alpha_{12}^{\prime}H_\mathrm{E}\left(\mathbf{s}\right) + \beta_{1}^{\prime}\delta_\mathrm{I}\left(\mathbf{s}\right),
\end{align}
where $\delta_\mathrm{I}\left(\mathbf{s}\right)$ is orthogonal to both $H_\mathrm{I}\left(\mathbf{s}\right)$ and $H_\mathrm{E}\left(\mathbf{s}\right)$, i.e., $\int_{\mathcal{S}_\mathrm{T}}H_\mathrm{I}^*\left(\mathbf{s}\right)\delta_\mathrm{I}\left(\mathbf{s}\right)d\mathbf{s}=0$ and $\int_{\mathcal{S}_\mathrm{T}}H_\mathrm{E}^*\left(\mathbf{s}\right)\delta_\mathrm{I}\left(\mathbf{s}\right)d\mathbf{s}=0$.
In this case, the secrecy rate can be expressed as given in~\eqref{eq:app1}, shown at the top of the next page.
\begin{figure*}[!h]
    \begin{align}
R_\mathrm{S}^{\prime} & = 
\operatorname{log}_2\left(1+\frac{\left|\int_{\mathcal{S}_\mathrm{T}}H_\mathrm{I}^*\left(\mathbf{s}\right)J_\mathrm{I}^{\prime}\left(\mathbf{s}\right)d\mathbf{s}\right|^2}{\left|\int_{\mathcal{S}_\mathrm{T}}H_\mathrm{I}^*\left(\mathbf{s}\right)J_\mathrm{A}\left(\mathbf{s}\right)d\mathbf{s}\right|^2+\sigma_\mathrm{I}^2}\right)- \operatorname{log}_2\left(1+\frac{\left|\int_{\mathcal{S}_\mathrm{T}}H_\mathrm{E}^*\left(\mathbf{s}\right)J_\mathrm{I}^{\prime}\left(\mathbf{s}\right)d\mathbf{s}\right|^2}{\left|\int_{\mathcal{S}_\mathrm{T}}H_\mathrm{E}^*\left(\mathbf{s}\right)J_\mathrm{A}\left(\mathbf{s}\right)d\mathbf{s}\right|^2+\sigma_\mathrm{E}^2}\right)\nonumber\\
& = \operatorname{log}_2\left(1+\frac{\left|\int_{\mathcal{S}_\mathrm{T}}H_\mathrm{I}^*(\mathbf{s})\left(\alpha_{11}^{\prime}H_\mathrm{I}\left(\mathbf{s}\right) + \alpha_{12}^{\prime}H_\mathrm{E}\left(\mathbf{s}\right)\right)d\mathbf{s}\right|^2}{\left|\int_{\mathcal{S}_\mathrm{T}}H_\mathrm{I}^*\left(\mathbf{s}\right)J_\mathrm{A}\left(\mathbf{s}\right)d\mathbf{s}\right|^2+\sigma_\mathrm{I}^2}\right)- \operatorname{log}_2\left(1+\frac{\left|\int_{\mathcal{S}_\mathrm{T}}H_\mathrm{E}^*\left(\mathbf{s}\right)\left(\alpha_{11}^{\prime}H_\mathrm{I}(\mathbf{s}) + \alpha_{12}^{\prime}H_\mathrm{E}\left(\mathbf{s}\right)\right)d\mathbf{s}\right|^2}{\left|\int_{\mathcal{S}_\mathrm{T}}H_\mathrm{E}^*\left(\mathbf{s}\right)J_\mathrm{A}\left(\mathbf{s}\right)d\mathbf{s}\right|^2+\sigma_\mathrm{E}^2}\right).
    \label{eq:app1}
\end{align}
\hrulefill
\end{figure*}

Next, we construct 
$\bar{J}_{I}\left(\mathbf{s}\right) = \bar{\alpha}_{11}H_\mathrm{I}\left(\mathbf{s}\right) + \bar{\alpha}_{12}H_\mathrm{E}\left(\mathbf{s}\right)$ with $\bar{\alpha}_{11}=\alpha_{11}^{\prime}$ and $\bar{\alpha}_{12}=\alpha_{12}^{\prime}$. Then, the achievable secrecy rate with $\bar{J}_{I}\left(\mathbf{s}\right)$ is given by
\begin{align}
\label{eq:app2}
\bar{R}_\mathrm{S} & = \operatorname{log}_2\left(1+\frac{\left|\int_{\mathcal{S}_\mathrm{T}}H_\mathrm{I}^*\left(\mathbf{s}\right)\left(\bar{\alpha}_{11}H_\mathrm{I}\left(\mathbf{s}\right) + \bar{\alpha}_{12}H_\mathrm{E}\left(\mathbf{s}\right)\right)d\mathbf{s}\right|^2}{\left|\int_{\mathcal{S}_\mathrm{T}}H_\mathrm{I}^*\left(\mathbf{s}\right)J_\mathrm{A}\left(\mathbf{s}\right)d\mathbf{s}\right|^2+\sigma_\mathrm{I}^2}\right)\nonumber\\
& - \operatorname{log}_2\left(1+\frac{\left|\int_{\mathcal{S}_\mathrm{T}}H_\mathrm{E}^*\left(\mathbf{s}\right)\left(\bar{\alpha}_{11}H_\mathrm{I}(\mathbf{s}) + \bar{\alpha}_{12}H_\mathrm{E}\left(\mathbf{s}\right)\right)d\mathbf{s}\right|^2}{\left|\int_{\mathcal{S}_\mathrm{T}}H_\mathrm{E}^*\left(\mathbf{s}\right)J_\mathrm{A}\left(\mathbf{s}\right)d\mathbf{s}\right|^2+\sigma_\mathrm{E}^2}\right).
\end{align}
Note that the expressions of the secrecy rate given in~\eqref{eq:app1} and~\eqref{eq:app2} are exactly the same. 

Denote the power of ${J}_\mathrm{I}^{\prime}\left(\mathbf{s}\right)$ and $\bar{J}_\mathrm{I}\left(\mathbf{s}\right)$ as $P_{\text{I}}^{\prime}$ and $\bar{P}_{\text{I}}$, respectively. Then, we have
\begin{align}
 \bar{P}_{\text{I}}&\triangleq\int_{\mathcal{S}_\mathrm{T}}\left|\bar{J}_\mathrm{I}\left(\mathbf{s}\right)\right|^2d\mathbf{s}\nonumber\\
& = \int_{\mathcal{S}_\mathrm{T}}\left|\bar{\alpha}_{11}H_\mathrm{I}\left(\mathbf{s}\right) + \bar{\alpha}_{12}H_\mathrm{E}\left(\mathbf{s}\right)\right|^2d\mathbf{s}\nonumber\\
& = \int_{\mathcal{S}_\mathrm{T}}\left|\alpha_{11}^{\prime}H_\mathrm{I}\left(\mathbf{s}\right) + \alpha_{12}^{\prime}H_\mathrm{E}\left(\mathbf{s}\right)\right|^2d\mathbf{s}\nonumber\\
& < \int_{\mathcal{S}_\mathrm{T}}\left|J_\mathrm{I}^{\prime}\left(\mathbf{s}\right)\right|^2d\mathbf{s}\triangleq P^{\prime}_{\text{I}}.
\end{align}
Therefore, we can say that, $\bar{J}_\mathrm{I}\left(\mathbf{s}\right)$ can achieve the same secrecy rate compared to ${J}_\mathrm{I}^{\prime}\left(\mathbf{s}\right)$, while possessing less power, i.e., ${J}_\mathrm{I}^{\prime}\left(\mathbf{s}\right)$ is not the optimal solution of problem~\eqref{eq:optimization_problem}. We reach a contradiction. The proof of~\eqref{eq:linerJI} is completed.

The optimal expression of AN source current pattern in ~\eqref{eq:linerJA} can be proved following a similar approach, for which the details are omitted here.
\label{app:A}

\end{appendices}

\bibliographystyle{IEEEtran}
\bibliography{mybib}
\end{document}